\documentclass[aps,pra,twocolumn,floatfix]{revtex4-1}

\usepackage{amssymb}
\usepackage{graphicx}
\usepackage{dcolumn}
\usepackage{bm}
\usepackage{amsmath}
\usepackage{epsfig}
\usepackage{hyperref}
\usepackage{braket}
\usepackage{caption}
\usepackage{epstopdf}
\usepackage{subcaption}
\usepackage{dsfont}
\usepackage{color}
\DeclareMathOperator{\Tr}{Tr}

\begin{document}

\title{Thermodynamic indistinguishability and field state fingerprint of quantum optical amplifiers}

\author{Y. Perl$^{1}$}
\author{Y. B. Band$^{1,2}$}
\author{E. Boukobza$^{3,4}$}
\affiliation{$^1$Department of Physics, Ben-Gurion University of the Negev, Beer-Sheva 8410501, Israel}
\affiliation{$^2$Department of Chemistry, Department of Electro-Optics, and the Ilse Katz Center for Nano-Science, Ben-Gurion University of the Negev, Beer-Sheva 8410501, Israel}
\affiliation{$^3$School of Chemistry, Tel-Aviv University, Tel-Aviv 6997801, Israel}
\affiliation{$^4$Nuclear Research Center Negev, Israel}

\begin{abstract}

Dissipation tends to wash out dynamical features observed at early evolution times. In this paper we analyze a resonant
single--atom two--photon quantum optical amplifier both dynamically
and thermodynamically. A detailed thermodynamic balance shows that
the non--linear amplifier is thermodynamically equivalent to the linear
amplifier discussed in Ref. \cite{Boukobza_Tannor2006a}. However, by calculating the Wigner
quasi--probability distribution for various initial field states, we show
that unique quantum features in optical phase space, absent from the linear
amplifier, are maintained for extended times. These features are related to
the discrete nature of the two--photon matter--field interaction, and fingerprint the
initial field state at thermodynamic times.

\end{abstract}

\maketitle

\section{Introduction}\label{sec:Intro}
Quantum thermodynamics has attracted increasing attention in the past two decades. Alicki pioneered the partitioning of energy fluxes in classically driven open quantum systems into heat fluxes and power \cite{Alicki}. Using Alicki's formalism, Geva \& Kosloff analyzed quantum systems in the framework of heat engines and heat pumps \cite{kosloff1992a,kosloff1992b,kosloff1994,kosloff1995,kosloff1996,kosloff2002,kosloff2004,kosloff2006,kosloff2010,kosloff2014a}. These systems can operate in strokes \cite{kosloff1992a,kosloff1992b,kosloff2002,kosloff2004}, or continuously \cite{kosloff1994,kosloff1995,kosloff1996}. A thermodynamic framework similar to that of Alicki but which incorporates a different dissipative mechanism was presented in \cite{Allahverdyan2005}. Other formalisms for partitioning energy into work and heat but without explicitly resorting to system--reservoir interactions include that of Allahverdyan and coworkers \cite{Allahverdyan_Mahler2008,Allahverdyan_Mahler2010}, and that of Quan et al \cite{Quan2007,Quan2009}.

In all of the above thermodynamical studies that employ Alicki's semiclassical formalism, the reservoir is not quantized. A pioneering formalism of quantum heat engines with a quantized reservoir is presented in \cite{Boukobza_Tannor2006b}. This formalism was applied to the processes of quantum light amplification and attenuation, and compared with its semiclassical analogue (using Alicki's formalism or an alternative semiclassical formalism) in \cite{Boukobza_Tannor2006a,Boukobza_Tannor2008,Boukobza_Ritsch}. The fact that the work reservoir is second quantized, enables one to dynamically study the quantum state of the piston.

More recent work treats the limits of operation of quantum machines. In \cite{Boukobza_Ritsch} a blue-detuned three-level quantum optical amplifier is shown to break the Carnot limit without violating the second law. Limiting efficiencies of quantum Otto engines are discussed in \cite{kosloff2014b,kosloff2014c}, and the limits of refrigeration of quantum systems are considered in \cite{kosloff2012a,kosloff2012b}. Using the notion of passivity (in fact its inverse, non--passivity), which was originally introduced by Lenard \cite{Lenard}, Gelbwaser-Klimovsky et.~al compute the amount of extractable work from quantum--piston amplifiers analyzed as heat engines \cite{Gelbwaser_Alicki,Gelbwaser_Kurizki}. Niedenzu et.~al delineated the role of coherences in power enhanced quantum heat engines, and showed that they still bound by the Carnot limit \cite{NiedenzuPRE2015}.

In this work we thermodynamically analyze the non--linear amplifier, and study the phase space dynamics of the quantum piston. This amplifier is the two--photon analogue of the linear amplifier discussed in \cite{Boukobza_Tannor2006a}. Related work studied the two--photon, non--linear Jaynes--Cummings (JCM) Hamiltonian \cite{Gerry1988}, where the unitary propagator was analytically derived. Also, complex collapse-revival dynamics in Kerr-like media were studied in \cite{Joshi1992}. An exact solution of a two-photon JCM with a dynamical Stark shift is presented in \cite{Iwasawa1999}.

This paper is arranged as follows. Section II introduces the non--linear quantum optical amplifier, and presents the level structure, the two--photon JCM Hamiltonian, and dissipative dynamics. In Sec. III we thermodynamically analyze the non--linear amplifier at steady state, based on the thermodynamical bipartite formalism presented in \cite{Boukobza_Ritsch} and \cite{Boukobza_Tannor2006b}. We show that from a steady state thermodynamic efficiency stand point, a non--linear amplifier is equivalent to a linear quantum optical amplifier. Section IV presents a phase space study of the field developed in the cavity at extended thermodynamic times. By calculating the Husimi--Kano Q function, we show that an initial even Fock state evolves into a phase diffused coherent state, which is indistinguishable at long times from an initial Poisson distributed field state with an identical average photon number. However, calculation of the Wigner function reveals the difference in the evolved field states with distinct negative quantum amplitudes for the even (or odd) initial Fock state. Section V contains a conclusion and a comparison with other recent quantum thermodynamical studies.

\section{Three--level amplifiers}
\begin{figure}[hbt]
\centering
\includegraphics[width=0.4\textwidth]{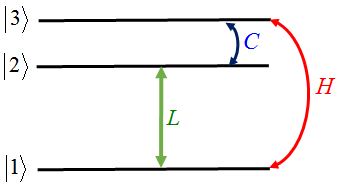}
\caption[Schematic diagram of a 3 level amplifier]{Diagram of a 3 level amplifier. A hot bath couples levels $\ket{1}$ and $\ket{3}$ and a cold bath couples levels $\ket{2}$ and $\ket{3}$. The matter--field interaction couples levels $\ket{1}$ and $\ket{2}$.}
\label{fig:LevelStruct}
\end{figure}

We analyze an amplifier similar to the one studied by Geva and Kosloff \cite{kosloff1994,kosloff1996} and by Boukobza and Tannor \cite{Boukobza_Tannor2006a,Boukobza_Tannor2007,Boukobza_Tannor2008}, with a different ordering of the atomic levels, more common in real optical amplifiers. The level structure is shown in Fig. \ref{fig:LevelStruct}. The hot bath couples levels $\ket{1}$ and $\ket{3}$ and the cold bath couples levels $\ket{2}$ and $\ket{3}$. The matter--field interaction couples levels $\ket{1}$ and $\ket{2}$. The main focus of this work is the non--linear amplifier, wherein two photons with a frequency of $\omega_f=\frac{\omega_2-\omega_1}{2}$ are emitted or absorbed by a single atomic transition. This non--linear version of the Jaynes Cummings model (JCM) was first introduced by Buck and Sukumar in 1981 \cite{Buck_Sukumar1981}, in the context of atom--phonon interactions.
The master equation consists of a two--photon atomic--field interaction Hamiltonian (unitary super-operator) and two atomic--bath (reservoir) Lindbladians (dissipative super-operators):
\begin{equation}\label{eq:Master}
\dot{\rho}=\mathcal{L}[\rho]=-\frac{i}{\hbar}[\mbox{\boldmath$H$},\mbox{\boldmath$\rho$}]+\mathcal{L}_{dH}[\mbox{\boldmath$\rho$}]+\mathcal{L}_{dC}[\mbox{\boldmath$\rho$}].
\end{equation}
The cold and hot dissipative parts of the Liouvillian are given in Lindblad \cite{Lindblad} form as:
\begin{equation}\label{eq:dis_liouvill}
\begin{split}
\mathcal{L}_{dH(C)}&[\mbox{\boldmath$\rho$}]=\\=&\Gamma_{H(C)}(\bar{n}_{H(C)}+1)
\left ( [\mbox{\boldmath$\sigma_{13(23)}\rho$},\mbox{\boldmath$\sigma_{13(23)}^{\dag}$}]
+H.C. \right )\nonumber\\
+&\bar{n}_{H(C)}\left ([\mbox{\boldmath$\sigma_{13(23)}^{\dag}\rho$},\mbox{\boldmath$\sigma_{13(23)}$}]
+H.C.)\right )
\end{split}
\end{equation}
Here $\bar{n}_{H(C)}$ is the thermal average of photons in the hot (cold) bath and is given by the Planck formula: $\bar{n}_{H(C)} = (\exp\left(\frac{\hbar\left(\omega_{3}-\omega_{1(2)}\right)}{k_BT_{H(C)}}\right)-1)^{-1}$. The $\boldsymbol{\sigma}$ matrices are the raising and lowering operators of the atom in the bipartite form, e.g., $\mbox{\boldmath$\sigma_{23}$}=\sigma_{23}\otimes\openone_{f}$, where $\sigma_{23}$ is the atomic lowering operator from level 3 to 2 and $\openone_{f}$ is the field identity operator. The Hamiltonian is given by:
\begin{equation}\label{eq:Hamil_complete}
\mbox{\boldmath$H$}=\mbox{\boldmath$H_{a}$}+\mbox{\boldmath$H_{f}$}+\mbox{\boldmath$H_{\mathrm{Int,2}}$},
\end{equation}
where $\mbox{\boldmath$H_{a}$}=H_{a}\otimes\openone_{f}$,
$\mbox{\boldmath$H_{f}$}=\openone_{a}\otimes H_{f}$, $H_{f}=\hbar\omega_{f}a^{\dag}a$ is the field Hamiltonian, $H_a=\hbar\sum_{i=1}^3\omega_i\ket{i}\bra{i}$ is the atomic Hamiltonian and $\openone_{a}$ is the atomic identity operator. The two--photon interaction rotating wave approximation (RWA) Hamiltonian is given by:
\begin{equation}\label{eq:Int_Hamil}
\mbox{\boldmath$H_{\mathrm{Int,2}}$}=\hbar\lambda[\sigma_{12} \otimes (a^{\dag})^2+\sigma_{12}^{\dag} \otimes a^2],
\end{equation}
where $a$($a^{\dag}$) is the lowering (raising) ladder operator of the field, and $\sigma_{12}$ ($\sigma_{12}^{\dag}$) is the lowering (raising) operator of levels 1 and 2 of the atom.
The dynamics of this two--photon Hamiltonian interaction of a two--level atom and a quantized field (without dissipation) was studied and solved analytically in \cite{Gerry1988}. As we will also conduct a comparative study with
the linear amplifier, we write the single--photon interaction Hamiltonian for completeness of presentation:
\begin{equation}\label{eq:Int1_Hamil}
\mbox{\boldmath$H_{\mathrm{Int}}$}=\lambda[\sigma_{12} \otimes (a^{\dag})+\sigma_{12}^{\dag} \otimes a].
\end{equation}
We note that for the linear amplifier case $\omega_{f}=\omega_{1}-\omega_{2}$ which is the resonance frequency, $\omega_{res}$.

\section{Thermodynamics of a non--linear amplifier}
The thermodynamic formalism we employ originates from the work of Boukobza and Tannor \cite{Boukobza_Tannor2006b}. This formalism applies to cases where the reservoir is also quantized, and is reminiscent of Alicki's formalism \cite{Alicki}. A slightly revised version of this formalism, which structurally captures better off--resonant excitations is obtained by expanding the time--derivative of the quasi--smiclassical energy operator
$\tilde{E}\equiv\Tr {\mbox{\boldmath$\rho \tilde{H}$}}$, where $\mbox{\boldmath$\tilde{H}$}=\mbox{\boldmath$H_a$}+\mbox{\boldmath$H_{\mathrm{Int,2}}$}$, whose time derivative is given by:
\begin{eqnarray}
\dot{\tilde{E}}&=&\Tr \left\{\mathcal{L}_{d}[\mbox{\boldmath$\rho$}](\mbox{\boldmath$H_A$}+\mbox{\boldmath$H_{\mathrm{Int,2}}$})\right\}+\frac{i}{\hbar}\Tr \left\{\mbox{\boldmath$\rho$}\left[\mbox{\boldmath$H_F$},\mbox{\boldmath$H_{\mathrm{Int,2}}$}\right] \right\}\nonumber\\&=&\dot{Q_H}+\dot{Q_C}-P_f
\end{eqnarray}
Here the thermodynamical currents of heat, $\dot{Q}_{H(C)}$, and field power, $P_{f}$, are defined similarly to \cite{Boukobza_Ritsch} as:
\begin{eqnarray}
\dot{Q}_{H(C)}&\equiv&\Tr \left\{\mathcal{L}_{dH(C)}[\mbox{\boldmath$\rho$}](\mbox{\boldmath$H_A$}+\mbox{\boldmath$H_{\mathrm{Int,2}}$})\right\} ,
\\
P_{f}&\equiv&-\frac{i}{\hbar}\Tr \left\{\mbox{\boldmath$\rho$}\left[\mbox{\boldmath$H_F$},\mbox{\boldmath$H_{\mathrm{Int,2}}$}\right] \right\} .
\end{eqnarray}

We solve the master equation \ref{eq:dis_liouvill} numerically, and study the long--time thermodynamic characteristics of the nonlinear amplifier in comparison with the linear amplifier presented in \cite{Boukobza_Tannor2006a,Boukobza_Tannor2006b}.
The parameters are chosen as follows: $\omega_{res}/\lambda=\lambda/\Gamma=10^3$, $(\omega_3-\omega_1)/\omega_{res}=1.2,\ (\omega_{3}-\omega_{2})/\omega_{res}=0.2$, $n_{H}=10$, $n_{C}=0.1$.

\begin{figure} [hbt!]
    \centering
    \begin{subfigure}[t] {0.2\textwidth}
        \includegraphics[width=\textwidth]
        {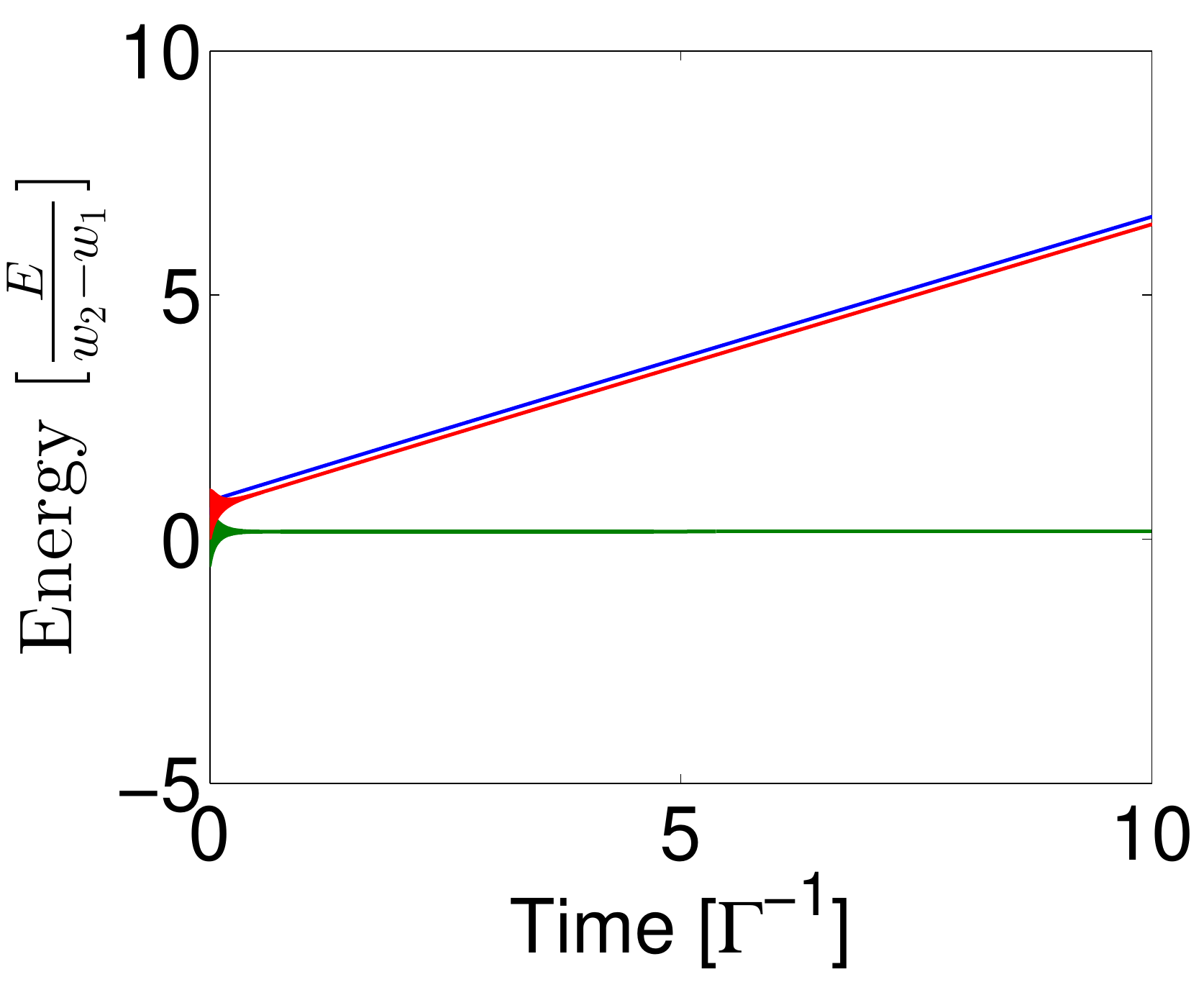}
        \vspace{-10pt}
    \caption{Energy}
     \label{fig:NonlinLongEnergy}
    \end{subfigure}
      \begin{subfigure}[t] {0.2\textwidth}
        \includegraphics[width=\textwidth]
        {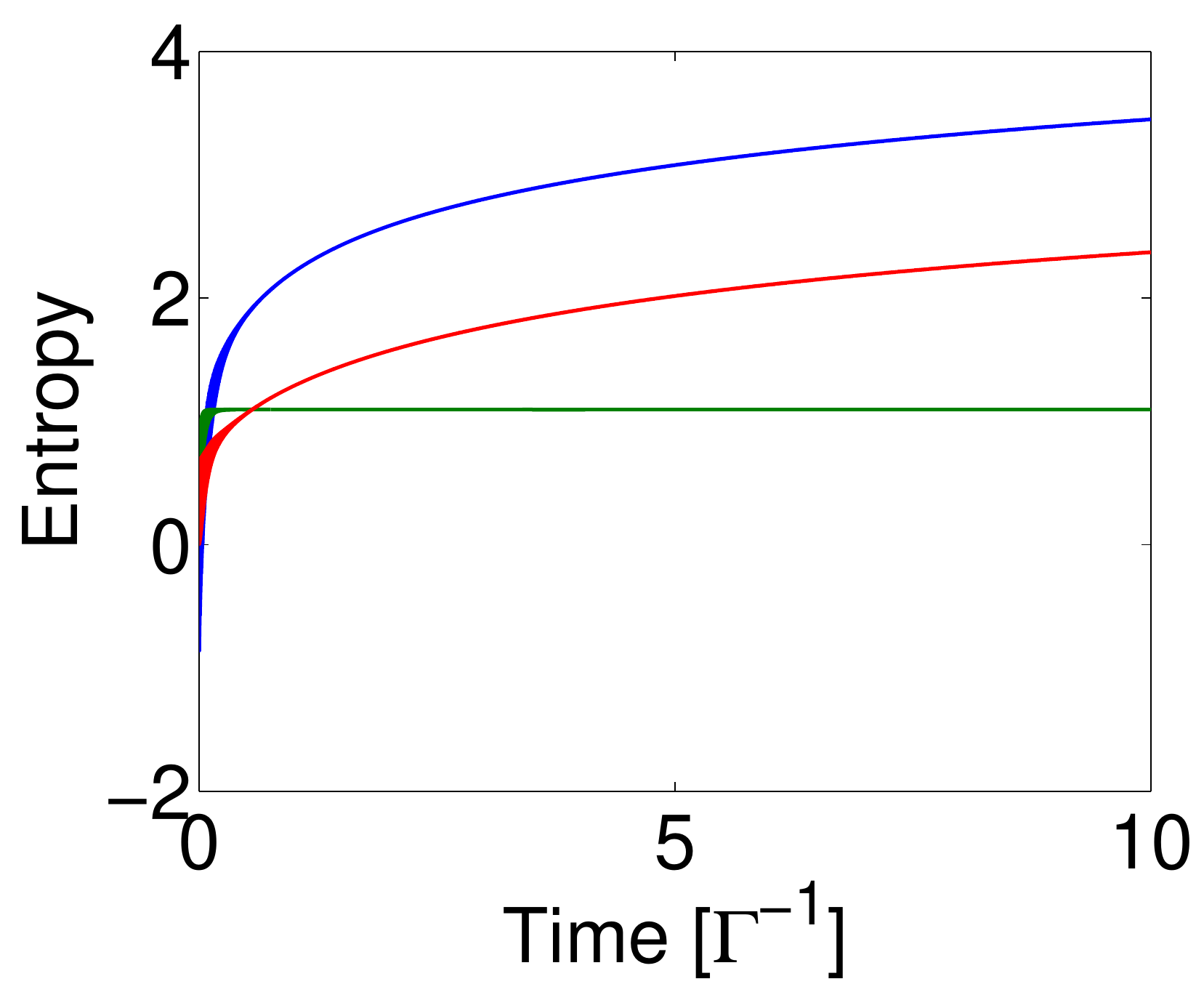}
      \vspace{-10pt}
    \caption{Entropy}
    \label{fig:NonlinLongEntropy}
    \end{subfigure}
        \begin{subfigure}[t] {0.2\textwidth}
        \includegraphics[width=\textwidth]
        {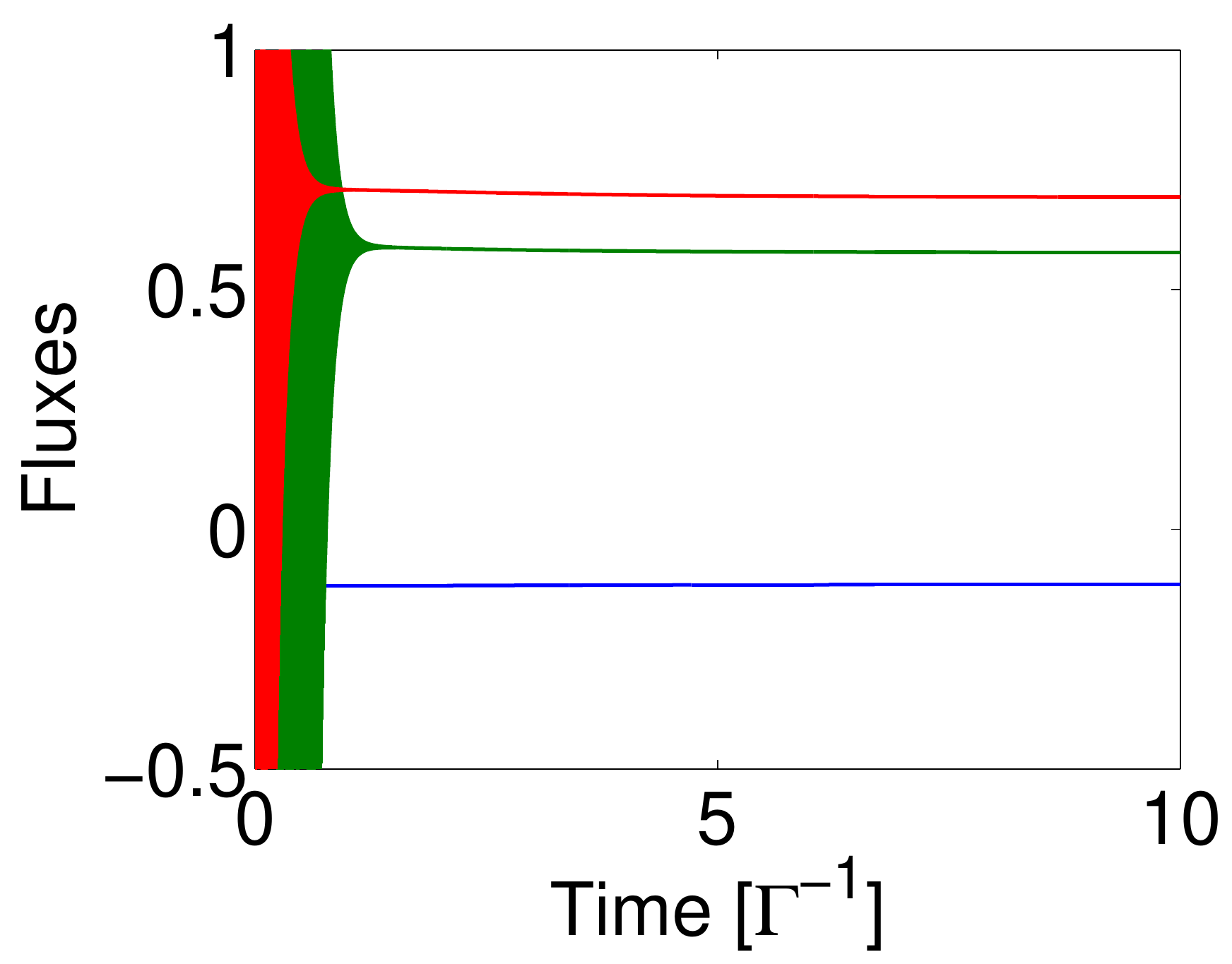}
         \vspace{-10pt}
    \caption{Heat Fluxes}
    \label{fig:NonlinLongHeatFlux}
    \end{subfigure}
        \begin{subfigure}[t] {0.2\textwidth}
        \includegraphics[width=\textwidth]
        {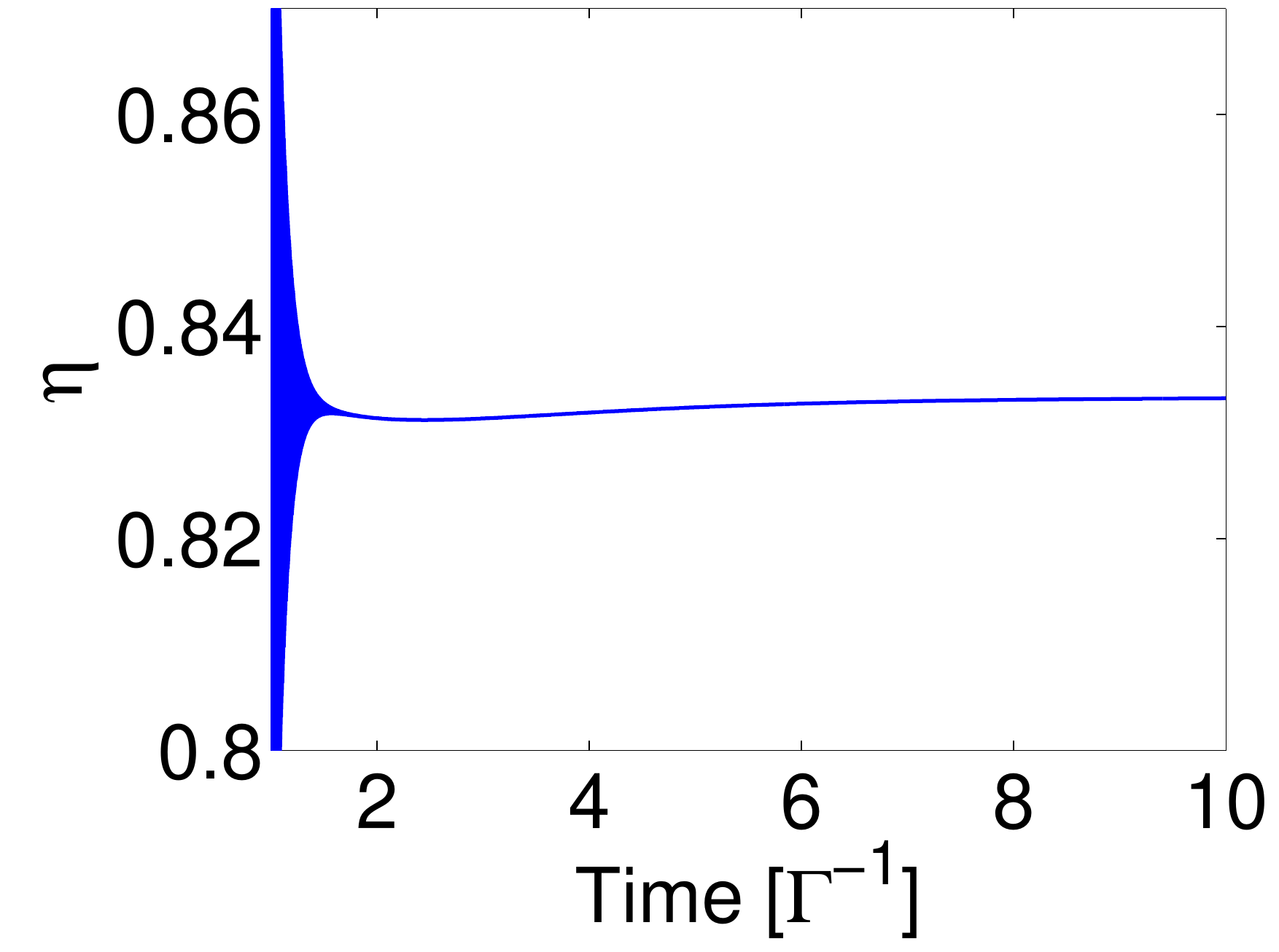}
        \vspace{-10pt}
    \caption{Efficiency}
     \label{fig:NonlinLongEff}
    \end{subfigure}
    \caption[Thermodynamic quantities vs. time for the non--linear amplifier]{Thermodynamical quantities vs. time for the non--linear amplifier. (a) Energy: atomic--field (blue), atomic (green), and field (red). (b) Entropy: atomic-field (blue), atomic (green), and field (red). (c) Energy currents: cold (blue) and hot (red) heat currents, and field power (green). (d) Efficiency.}
  \label{fig:NonlinLongThermo}
\end{figure}

To demonstrate that amplification occurs even for an empty cavity, we choose a separable initial state, where the atom is excited and the field is in the zero photon Fock state (empty cavity).
The master equation is solved on--resonance with the fourth order Runge--Kutta method until $t=10 \, \Gamma^{-1}$. Figure \ref{fig:NonlinLongThermo} plots the evolution in time of several thermodynamical quantities. Figure \ref{fig:NonlinLongEnergy} shows the energy of each sub--system (atomic green, field red) as well as the total energy of the system (blue). Figure \ref{fig:NonlinLongEntropy} shows the entropy of each sub--system (atomic green, field red) as well as the total entropy of the system (blue). As in the linear amplifier, the atomic sub--system reaches steady state while the field sub--system continues to grow in both energy and entropy, indicating a non--internal coherence amplification process as noted in \cite{Boukobza_Ritsch}. Figure \ref{fig:NonlinLongHeatFlux} shows the cold and hot heat fluxes, as well as the power. The steady-state values of the fluxes are: $\dot{Q}_H=0.6927$, $\dot{Q}_C=-0.1155$ and $P_f=0.5773$, indicating that the non--linear quantum optical amplifier is thermodynamically equivalent to a heat engine (heat is extracted from the hot reservoir, and in turn is simultaneously channeled to optical output power and dissipated to the cold reservoir. These results match those obtained for the linear amplifier with an identical choice of coupling and optical parameters.
Figure \ref{fig:NonlinLongEff} shows the efficiency of the amplifier, defined naturally as $\eta=\frac{P_{f}}{\dot{Q}_{H}}$. The efficiency reaches the value predicted for a linear amplifier by Scovil and Schulz-DuBois \cite{scov_schulz_dubois}, $\eta=\frac{\omega_{2}-\omega_{1}}{\omega_{3}-\omega_{1}} = 0.8333$.

\begin{figure} [htb]
    \centering
    \begin{subfigure}[h] {0.4\textwidth}
        \includegraphics[width=\textwidth]
        {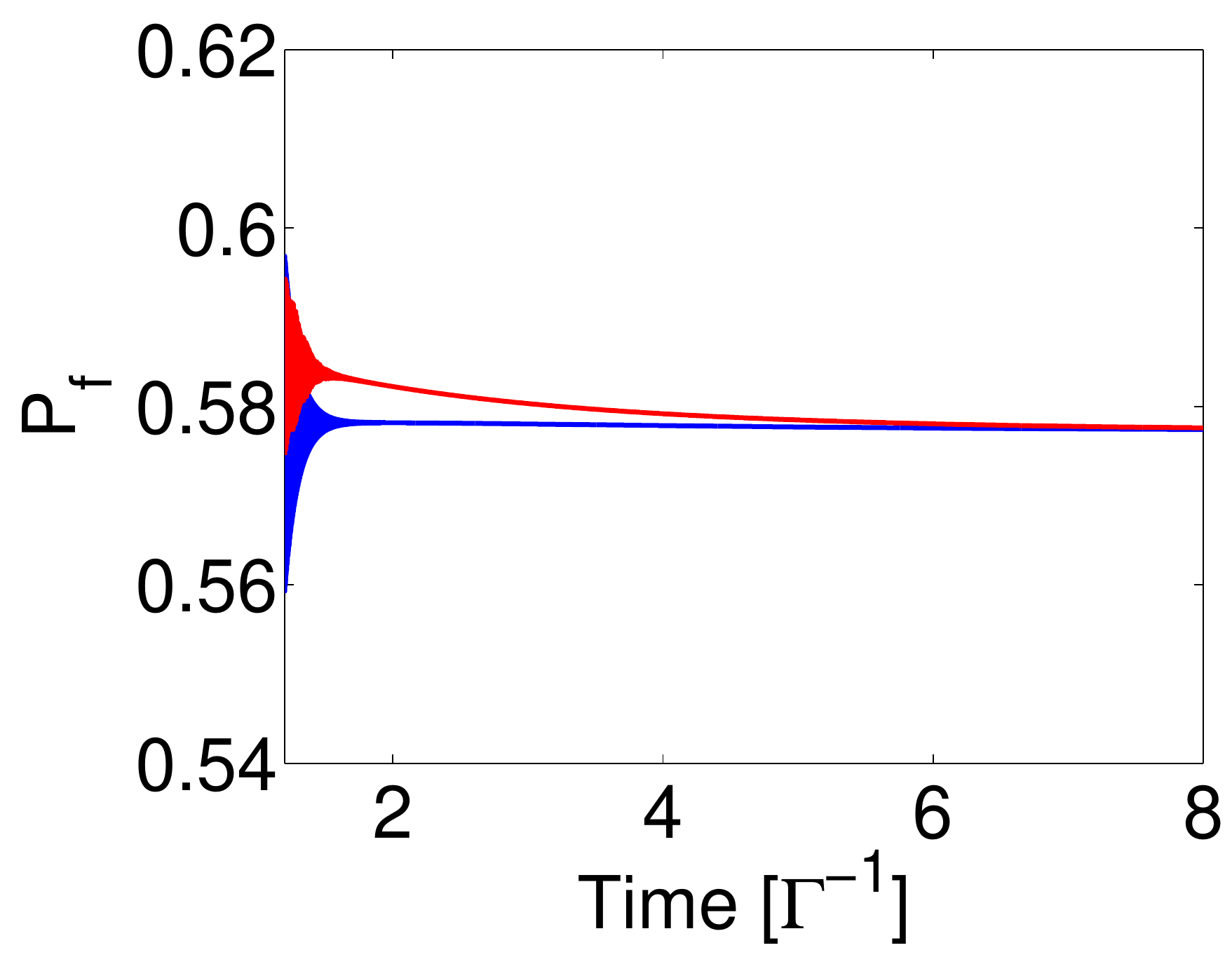}
    \caption{Power}
    \label{fig:ComparePf}
    \end{subfigure}
        \begin{subfigure}[h] {0.4\textwidth}
        \includegraphics[width=\textwidth]
        {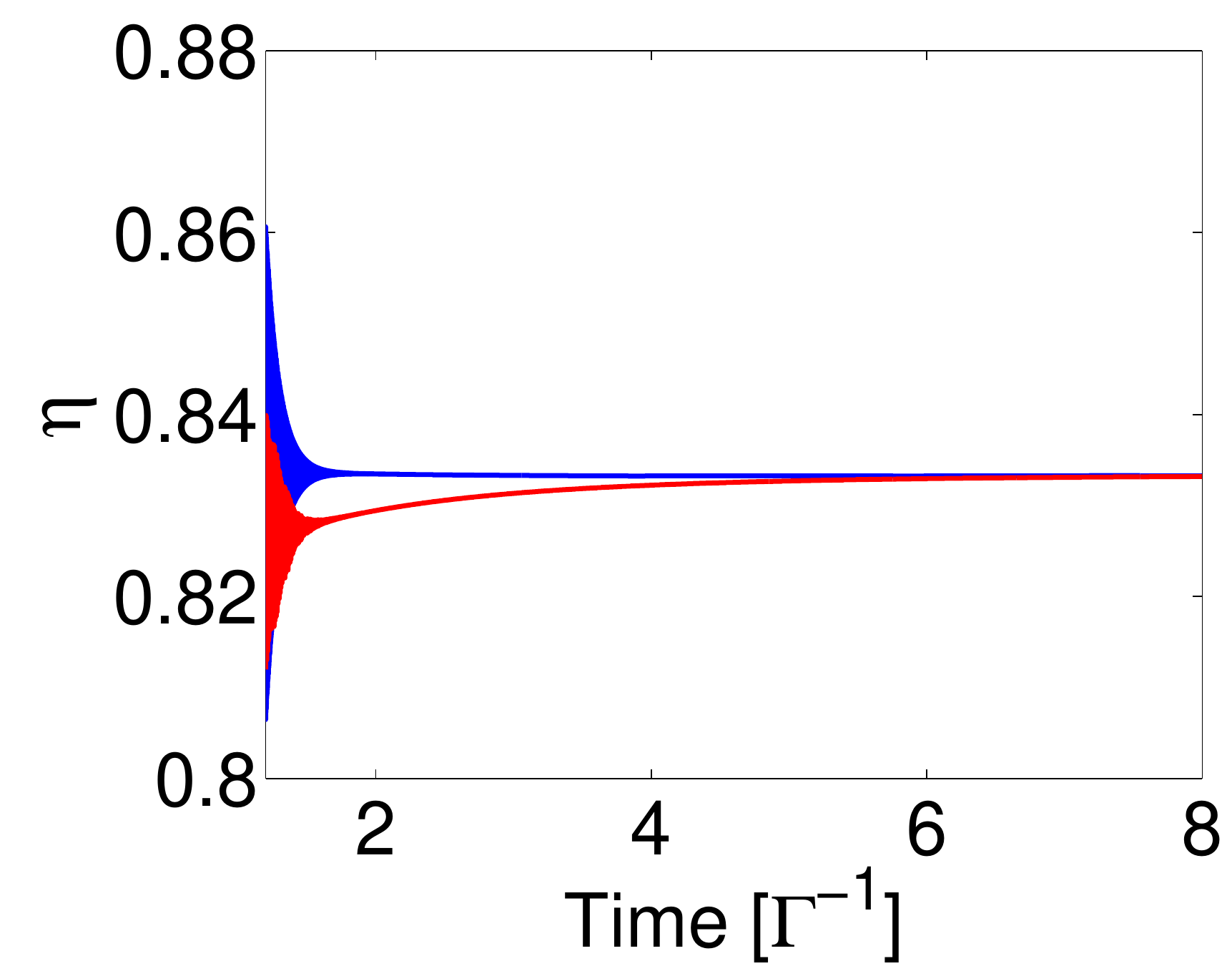}
    \caption{Efficiency}
    \label{fig:CompareEta}
    \end{subfigure}
  \caption[Power and efficiency in the non--linear amplifier]{Power and efficiency in the non--linear amplifier: (a) Power and (b) efficiency, for an initial Fock state (blue) and an initial mixed state (red).}
       \label{fig:ComparePfEta}
\end{figure}

In order to determine whether the thermodynamic features of the non--linear amplifier are generic regardless of the initial state, we solve the master equation for extended times for two initial field states. The atom is initially in the ground state, while the field is initially either in a Fock state with 4 photons, i.e., $\rho_{f}\left(t=0\right)=\ket{4}\bra{4}$, or in a mixed state with Poisson photon distribution and mean photon number of 4, i.e., $\rho_{f}\left(t=0\right)=\sum_{n=0}^{\infty}\frac{e^{-4}4^n}{n!}\ket{n}\bra{n}$. The latter field state has an identical population occupation as a coherent state with the same mean photon number, but has no well--defined phase (no internal coherence in the Fock representation). Figure \ref{fig:ComparePf} shows the steady state field power and Fig.~\ref{fig:CompareEta} plots the efficiency. We can see that the thermodynamical quantities at long times converge to the same value for the two states, hence, they are thermodynamically equivalent. We note that this observation holds also for the individual heat current components.

Before analyzing various phase space representations of the field state, we will present here a short straightforward derivation of the efficiency formula, based on a thermodynamical analysis of the semiclassical non--linear amplifier. The master equation in the semiclassical regime is similar to Eq. \ref{eq:Master}. We emphasize that the atomic dissipative super--operators are practically identical to the fully quantized case (but they exist in a $C^3$ Hilbert space). Secondly, the atomic-field second--quantized Hamiltonian is replaced by the semiclassical RWA two--photon coupling Hamiltonian:
\begin{equation}
H=H_{a}+H_{Int,2}^{sc};\ \ H_{Int,2}^{sc}= \hbar\lambda[\sigma_{12}e^{2i\omega_{f}}+\sigma_{12}^{\dag}e^{-2i\omega_{f}}].
\end{equation}
The semiclassical master equation can be solved at steady state in the (two--photon) rotating frame (this follows the same line of derivation as appears in \cite{Boukobza_Tannor2007}). Alternatively, a steady state value for the average values of the energy/entropy might be sought in the Schr\"{o}dinger picture. Nevertheless, the steady state values for the hot and cold heat currents as well as the power are given by ($\dot{E}^{sc}=0$):
\begin{equation}
\begin{split}
&\dot{Q}_{H}^{sc}\equiv\Tr{ \left \{ \mathcal{L}_{dH}[\rho]\cdot H_a \right \} }=\\
&2\Gamma_H\Gamma_C\lambda^2(n_H - n_C)(\omega_3 - \omega_1)\frac{\Gamma_Hn_H + \Gamma_Cn_C}{\beta\lambda^2+\gamma}
\end{split}
\end{equation}
\begin{equation}
\begin{split}
&\dot{Q}_{C}^{sc}\equiv\Tr{\left \{\mathcal{L}_{dC}[\rho]\cdot H_a\right \}}=\\
&-2\Gamma_H\Gamma_C\lambda^2(n_H - n_C)(\omega_3 - \omega_2)\frac{\Gamma_Hn_H + \Gamma_Cn_C}{\beta\lambda^2+\gamma}
\end{split}
\end{equation}
\begin{equation}
\begin{split}
&P^{sc}\equiv-\Tr{\left \{\rho\cdot\frac{dV(t)}{dt} \right \}}=\\
&2\Gamma_H\Gamma_C(n_H - n_C)(\Gamma_Hn_H + \Gamma_Cn_C)\frac{(\omega_2 - \omega_1)}{\beta\lambda^2+\gamma}\lambda^2
\end{split}
\end{equation}
where the $\equiv$ indicates the definition for the heat currents and power for externally driven systems as originally conceived by Alicki \cite{Alicki}, and for convenience we have used $\alpha\equiv\Gamma_H\Gamma_C(n_C+n_H+ 3n_Hn_C)$, $\beta=(\Gamma_Hn_H + \Gamma_Cn_C)(2\Gamma_H + 2\Gamma_C + 3\Gamma_Hn_H + 3\Gamma_Cn_C)$ and $\gamma=\alpha(\Gamma_Hn_H + \Gamma_Cn_C)^2$. Dividing $P^{sc}$ by $\dot{Q_{H}^{sc}}$ yields the amplifier efficiency:
\begin{equation}
\eta=\frac{\omega_{2}-\omega_{1}}{\omega_{3}-\omega_{1}}=\frac{\omega_{res}}{\omega_{p}},
\end{equation}
where $\omega_{p}=\omega_{3}-\omega_{1}$ is the (central) pumping frequency. The efficiency of the two--photon resonant amplifier is identical to that of the resonant linear amplifier discussed in \cite{Boukobza_Tannor2006a}. This suggests that an efficiency formula for resonant multi--photonic optical amplifiers, which is given as a ratio of the resonance frequency to the pump frequency, might be generic regardless of the photonic cascade. However, in deriving such a result by way of induction, one should take care that $\omega_{\mathrm{res}}/n$ is not close to any of the central reservoir frequency, as the validity of the master equation would then be questionable.

\section{Phase space representations of quantum optical amplifiers}\label{sec:Phase_space}
In the previous section we established that two different initial field states are thermodynamically indistinguishable at long times. In this section we determine weather two different initial photonic states have unique characterizing features in optical phase space at long times. We base our optical phase space analysis on two functions.
The first is the Hussimi--Kano Q--function, introduced by Husimi in 1940 \cite{Husimi1940}, and is defined by:
\begin{equation}\label{eq:HussimiKanoDef}
Q\left(\alpha\right)\equiv\frac{1}{\pi}\bra{\alpha}\rho_f\ket{\alpha},
\end{equation}
where $\ket{\alpha}$ is the coherent state:\\ $\ket{\alpha}=e^{-\frac{\left|\alpha\right|^2}{2}}\sum_{n=0}^{\infty}\frac{\alpha^n}{\sqrt{n!}}\ket{n}$.

The second phase space function, which will be of more interest to us, is the Wigner distribution function, introduced by Wigner in 1932 \cite{Wigner1932}, and is defined by:
\begin{equation}\label{eq:wigner_def1}
W\left(x,p\right)\equiv\frac{1}{\pi\hbar}\int_{-\infty}^{\infty}{\bra{x+y}\rho_f\ket{x-y}e^{-2i py/\hbar}dy}.
\end{equation}

\begin{figure} [htb]
    \centering
    \begin{subfigure}[h] {0.12\textwidth}
        \includegraphics[width=\textwidth]
        {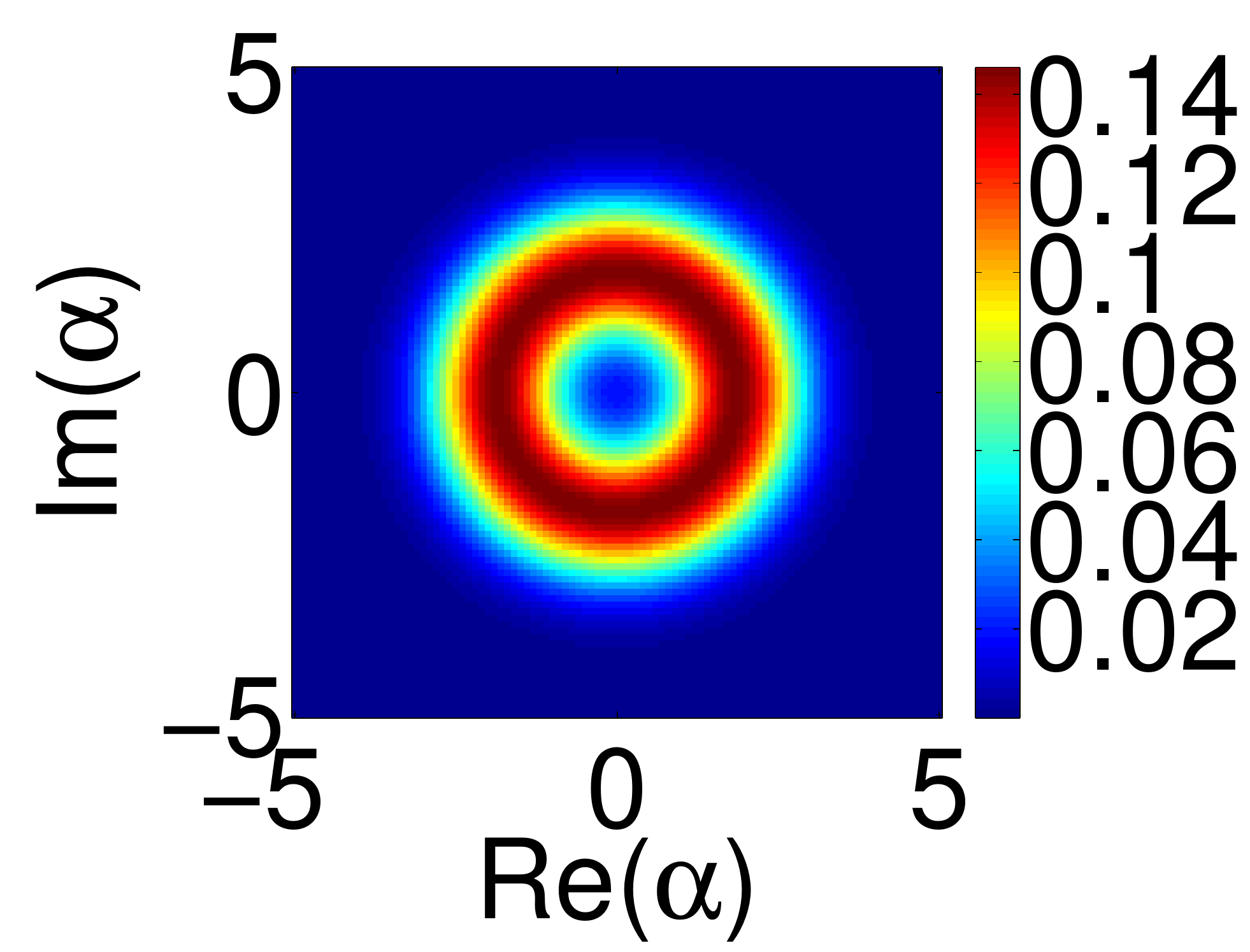}
    \caption{Mixed, $t=0$}
    \label{fig:QMixed0}
    \end{subfigure}
        \begin{subfigure}[h] {0.12\textwidth}
        \includegraphics[width=\textwidth]
        {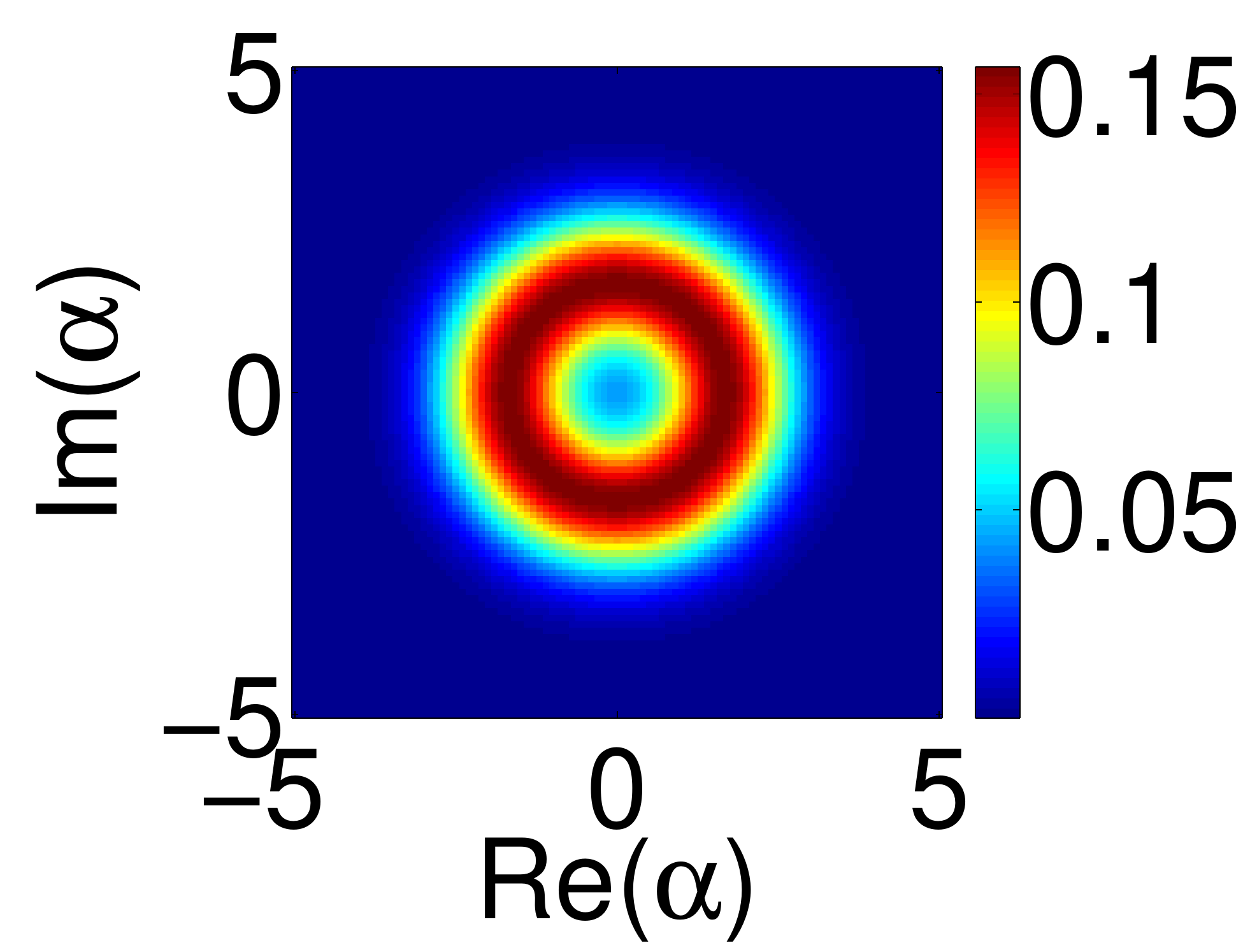}
    \caption{Mixed, $t=0.1 \, \Gamma^{-1}$}
    \label{fig:QMixed01}
    \end{subfigure}
        \begin{subfigure}[h] {0.12\textwidth}
        \includegraphics[width=\textwidth]
        {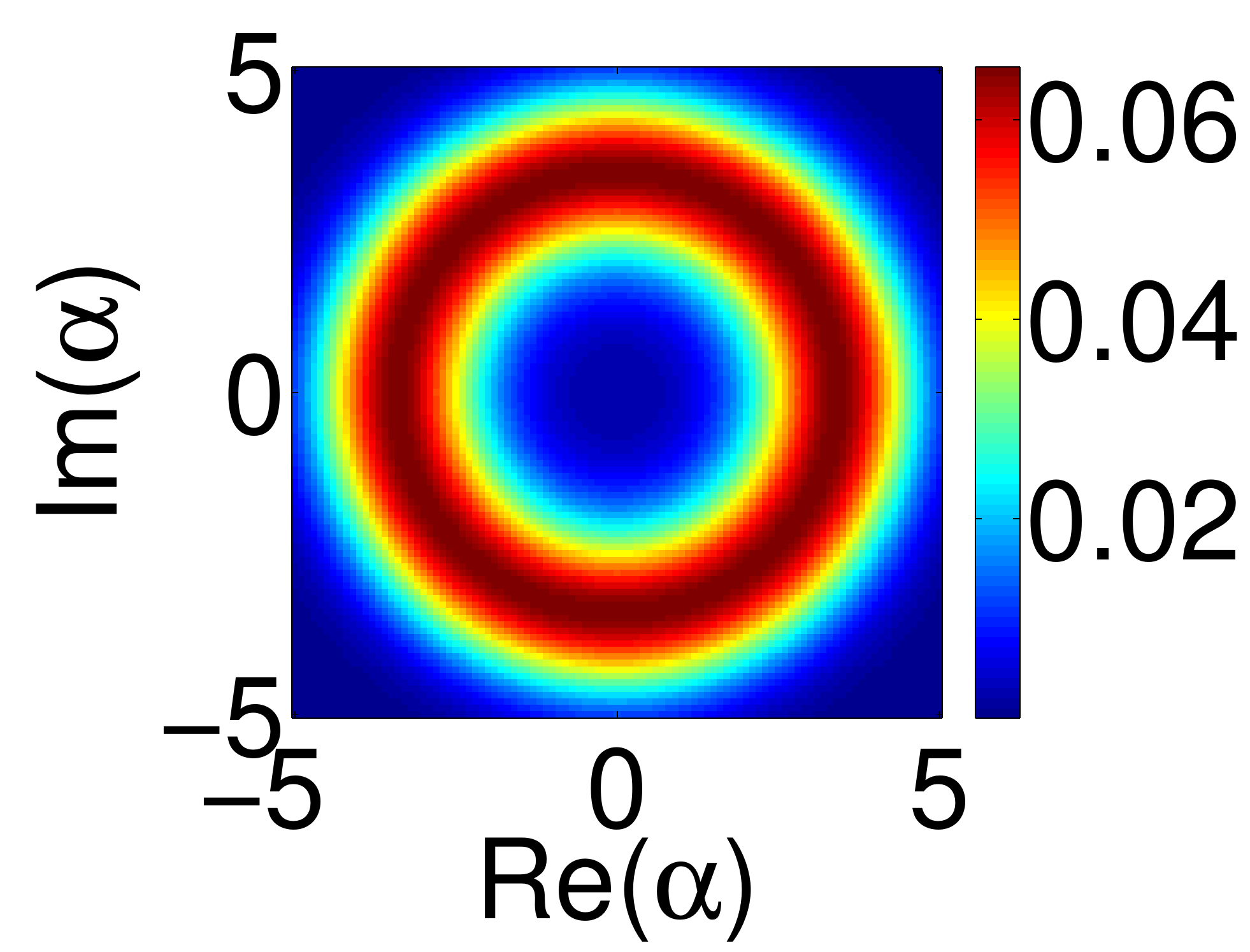}
    \caption{Mixed, $t=8 \, \Gamma^{-1}$}
    \label{fig:QMixed8}
       \end{subfigure}
          \begin{subfigure}[h] {0.12\textwidth} \vspace{-10pt}
        \includegraphics[width=\textwidth]
        {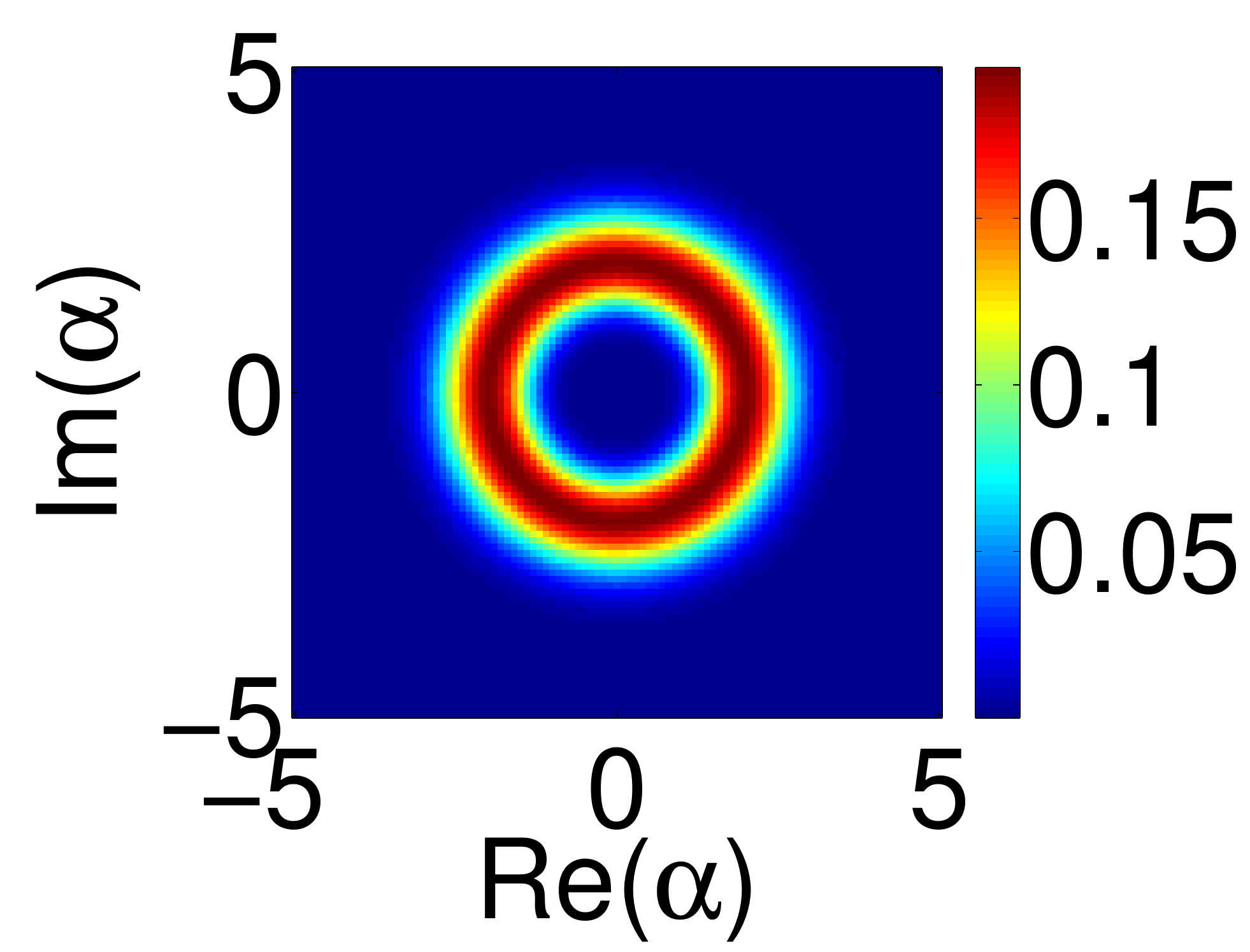}
    \caption{Fock, $t=0$}
    \label{fig:QFock0}
    \end{subfigure}
        \begin{subfigure}[h] {0.12\textwidth}
        \includegraphics[width=\textwidth]
        {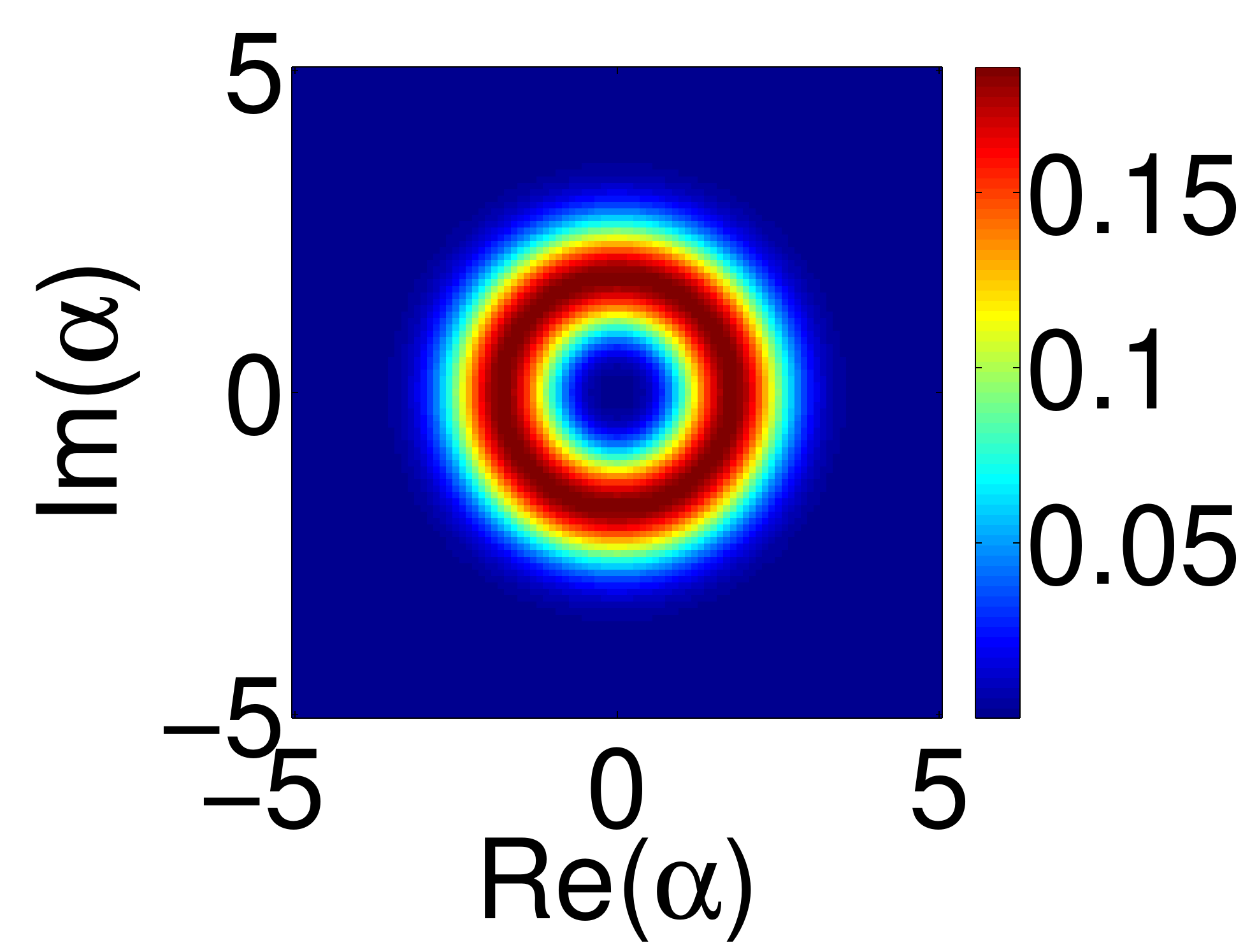}
    \caption{Fock, $t=0.1 \, \Gamma^{-1}$}
    \label{fig:QFock01}
    \end{subfigure}
        \begin{subfigure}[h] {0.12\textwidth}
        \includegraphics[width=\textwidth]
        {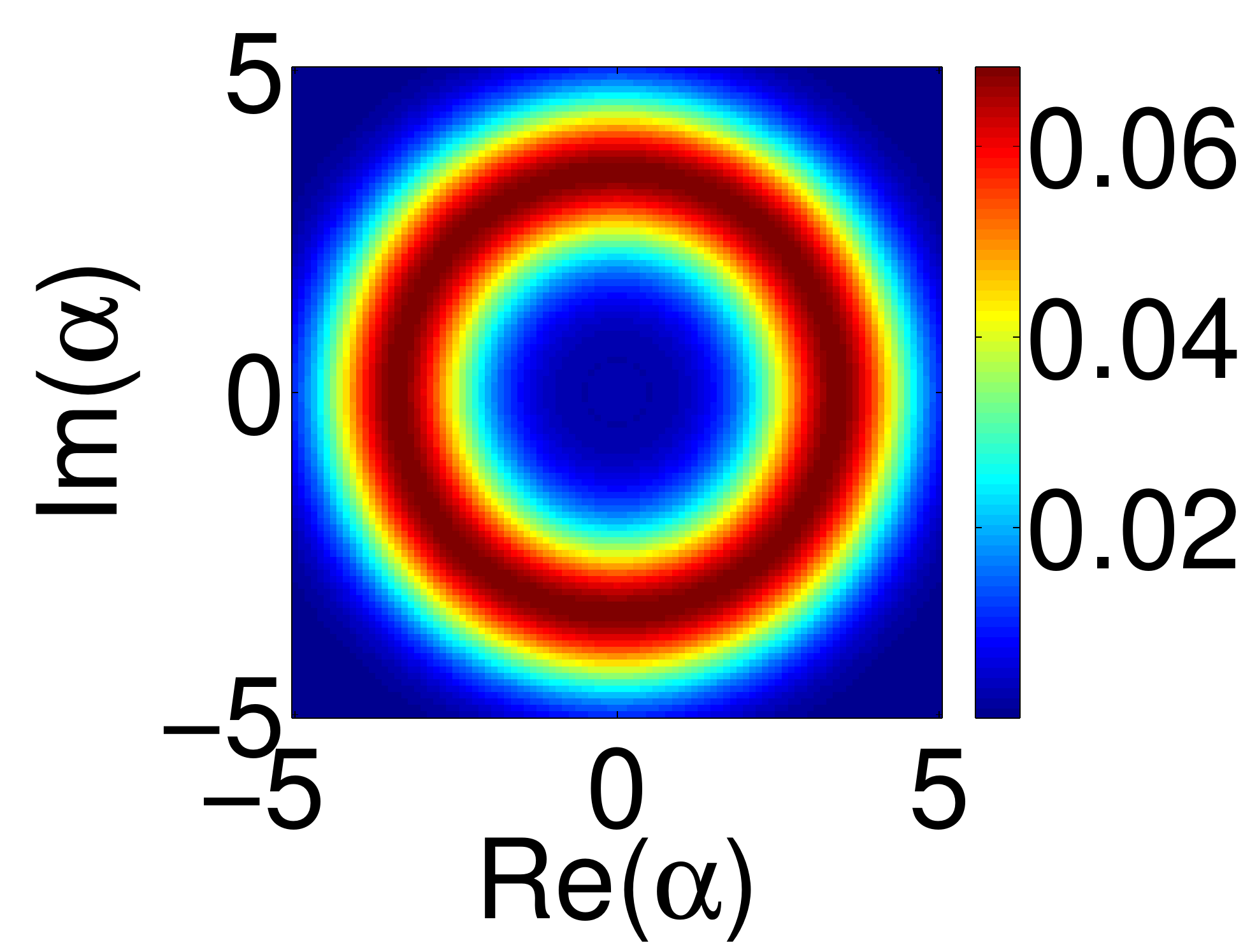}
    \caption{Fock, $t=8 \, \Gamma^{-1}$}
    \label{fig:QFock8}
       \end{subfigure}
\caption[Non linear interaction: Hussimi--Kano Q--function at various times]{Two--photon amplifier Hussimi--Kano Q--functions at various times for an initial Poissonian--mixed state with $\bar{n}=4$ (a-c) and for an initial $n=4$ Fock state (d-f)}
       \label{fig:QFun}
\end{figure}

We begin by plotting the Hussimi--Kano Q--function at different times for both the initial diagonal Poissonian (mixed) state with $\bar{n}=4$ in Fig. \ref{fig:QFun}(a)-(c), and the $n=4$ Fock (pure) state in Fig. \ref{fig:QFun}(d)-(f) (the same states considered in Fig.~\ref{fig:ComparePfEta}). We see that although the Q--functions for the two states are somewhat different initially, after $8 \, \Gamma^{-1}$ they become practically indistinguishable (the increased radius of the cylindrically symmetric quasiprobability reflects field amplification). We note that the colors used to plot the Q--functions are normalized to display each image at maximal color depth.

\begin{figure} [htb]
    \centering
    \begin{subfigure}[h] {0.12\textwidth}
        \includegraphics[width=\textwidth]
        {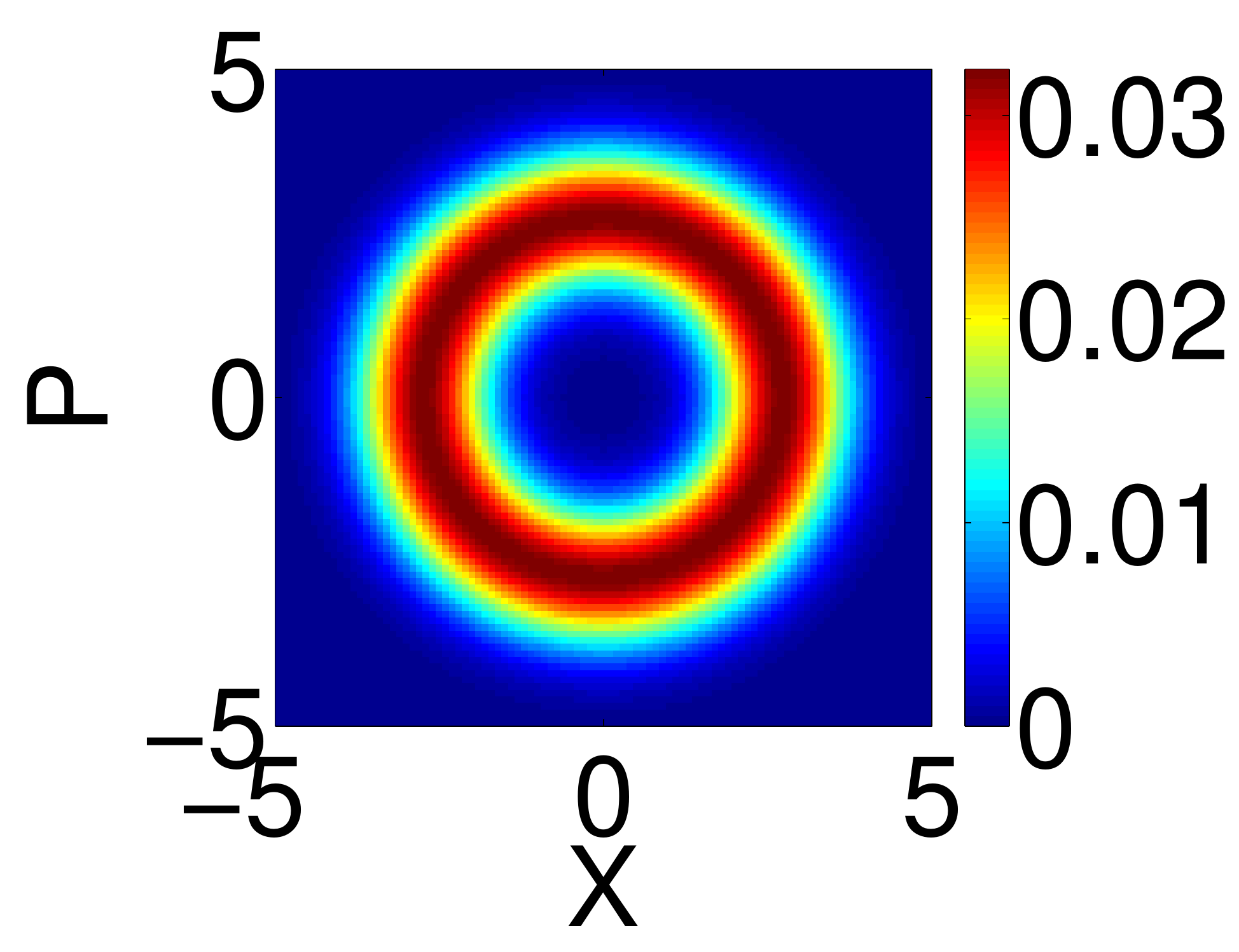}
    \caption{Mixed, $t=0$}
    \label{fig:WMixed0}
    \end{subfigure}
        \begin{subfigure}[h] {0.12\textwidth}
        \includegraphics[width=\textwidth]
        {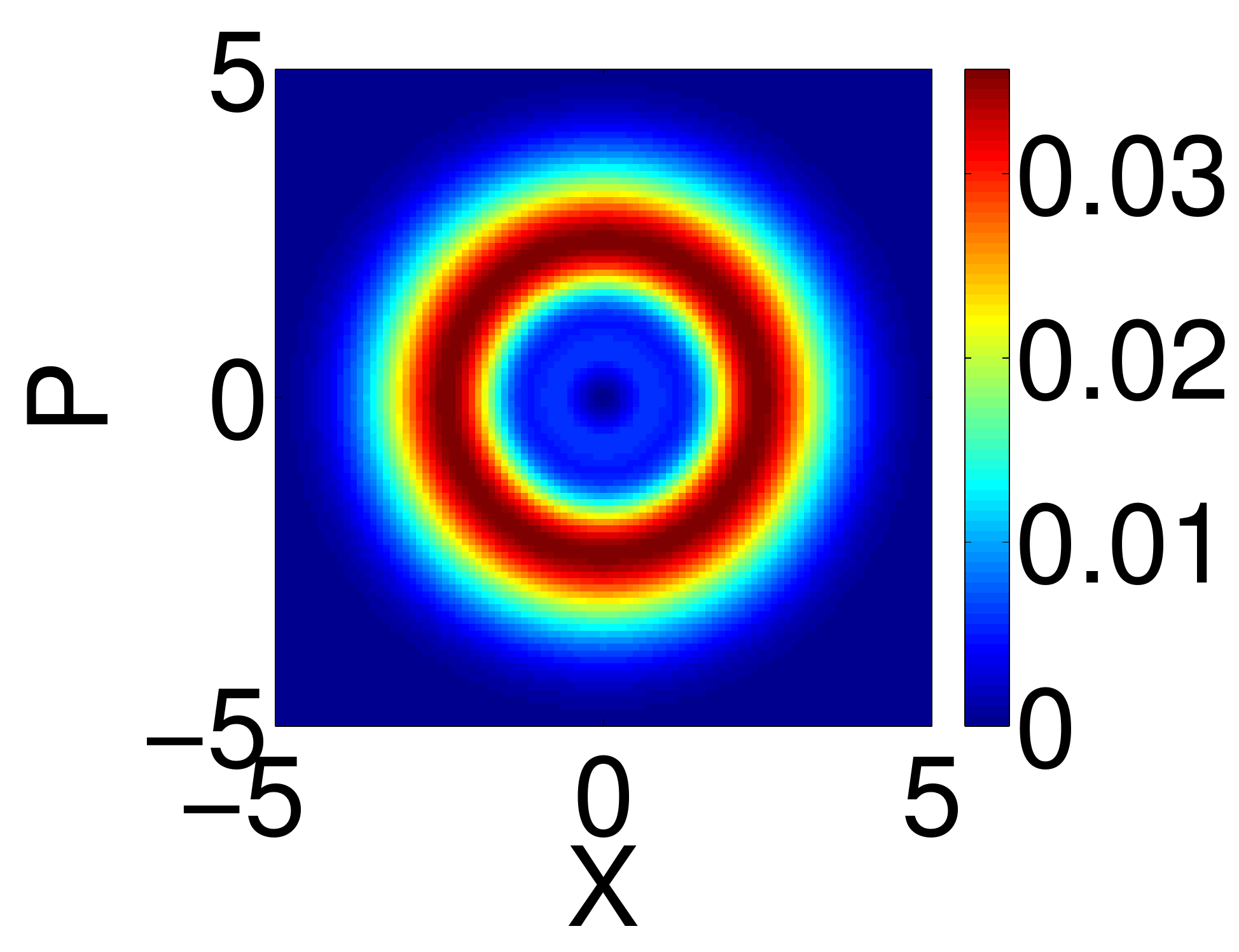}
    \caption{Mixed, $t=0.1 \, \Gamma^{-1}$}
    \label{fig:WMixed01}
    \end{subfigure}
        \begin{subfigure}[h] {0.12\textwidth}
        \includegraphics[width=\textwidth]
        {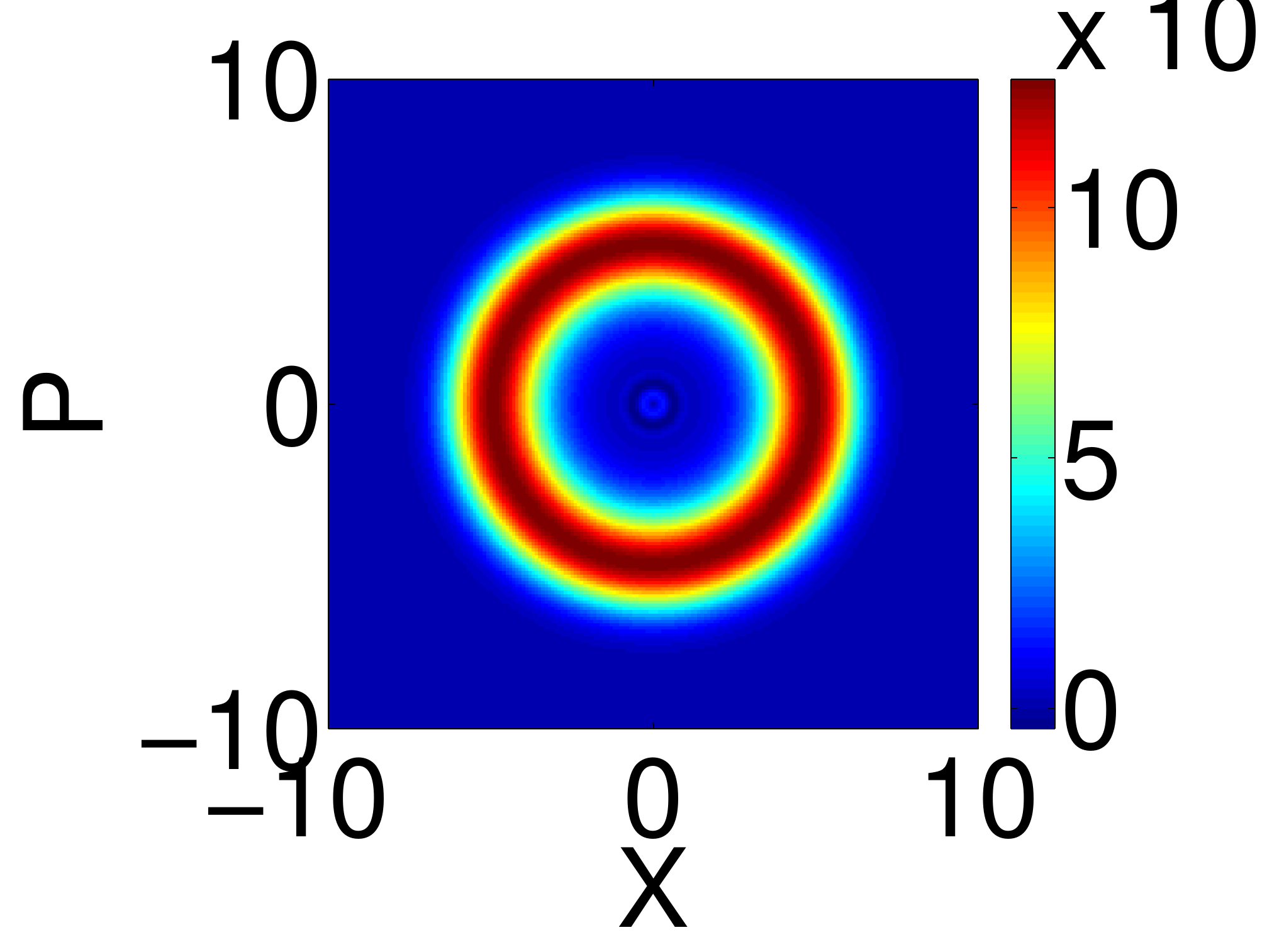}
    \caption{Mixed, $t=8 \, \Gamma^{-1}$}
    \label{fig:WMixed8}
       \end{subfigure}
          \begin{subfigure}[h] {0.12\textwidth} \vspace{-10pt}
        \includegraphics[width=\textwidth]
        {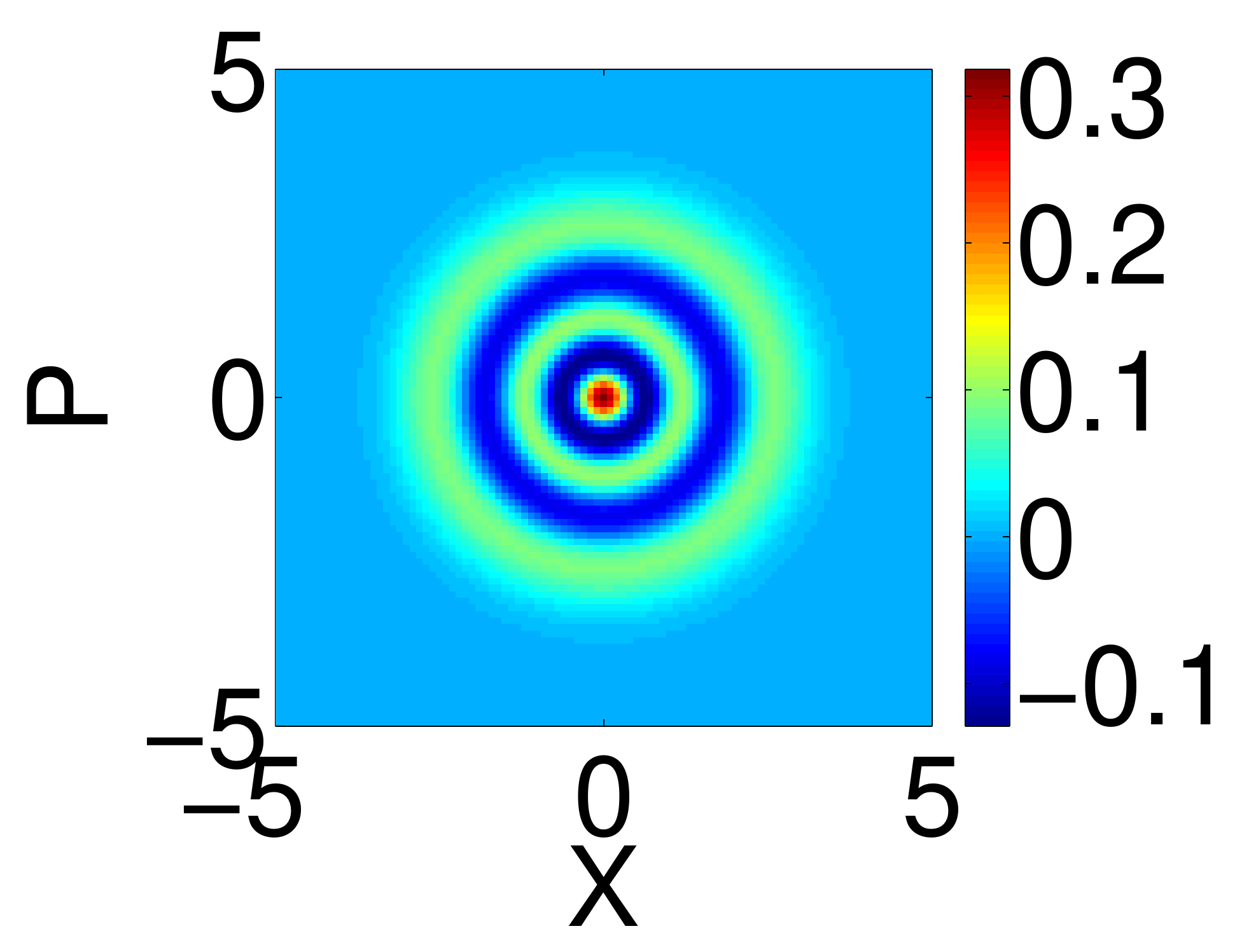}
    \caption{Fock, $t=0$}
    \label{fig:WFock0}
    \end{subfigure}
        \begin{subfigure}[h] {0.12\textwidth}
        \includegraphics[width=\textwidth]
        {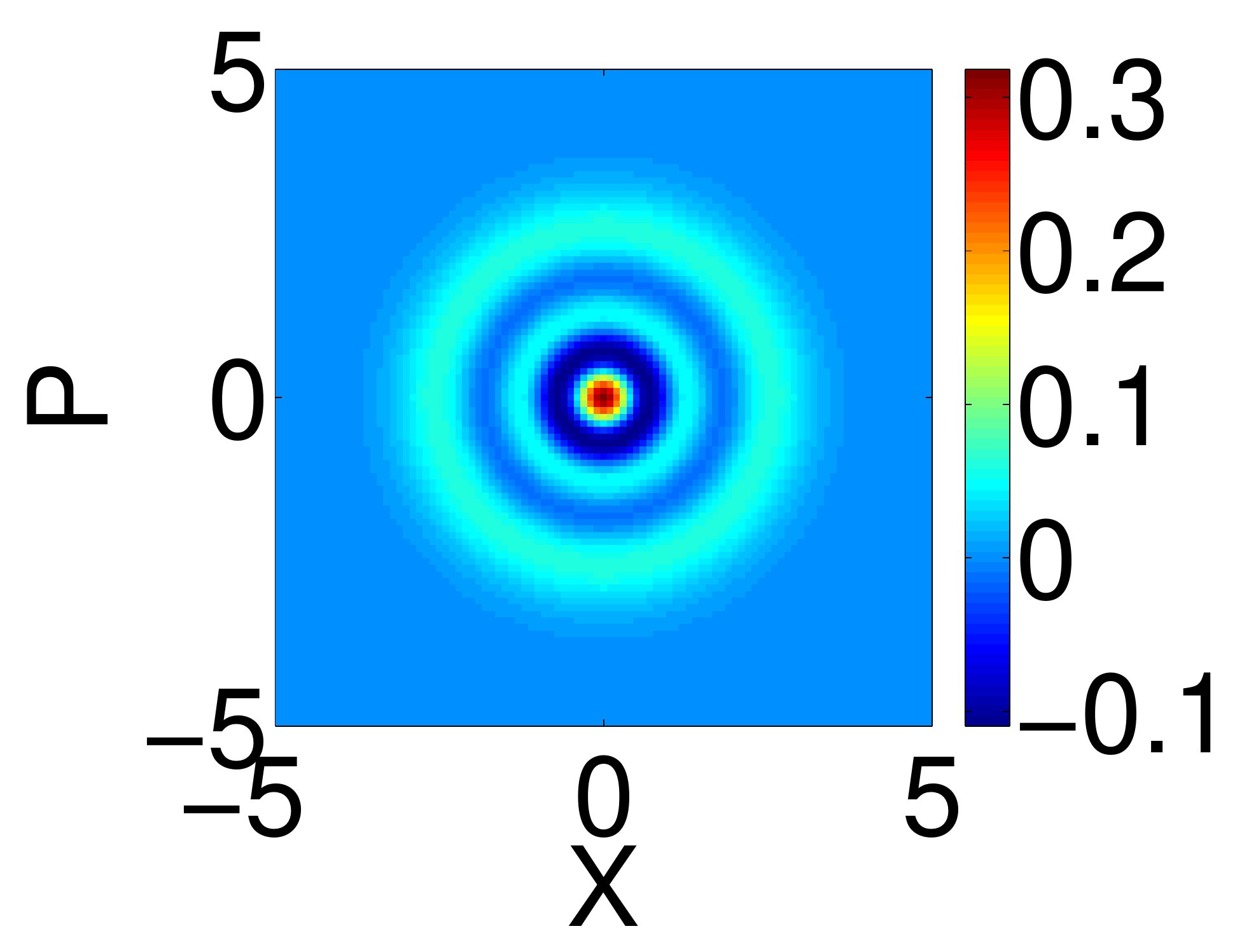}
    \caption{Fock, $t=0.1 \, \Gamma^{-1}$}
    \label{fig:WFock01}
    \end{subfigure}
        \begin{subfigure}[h] {0.12\textwidth}
        \includegraphics[width=\textwidth]
        {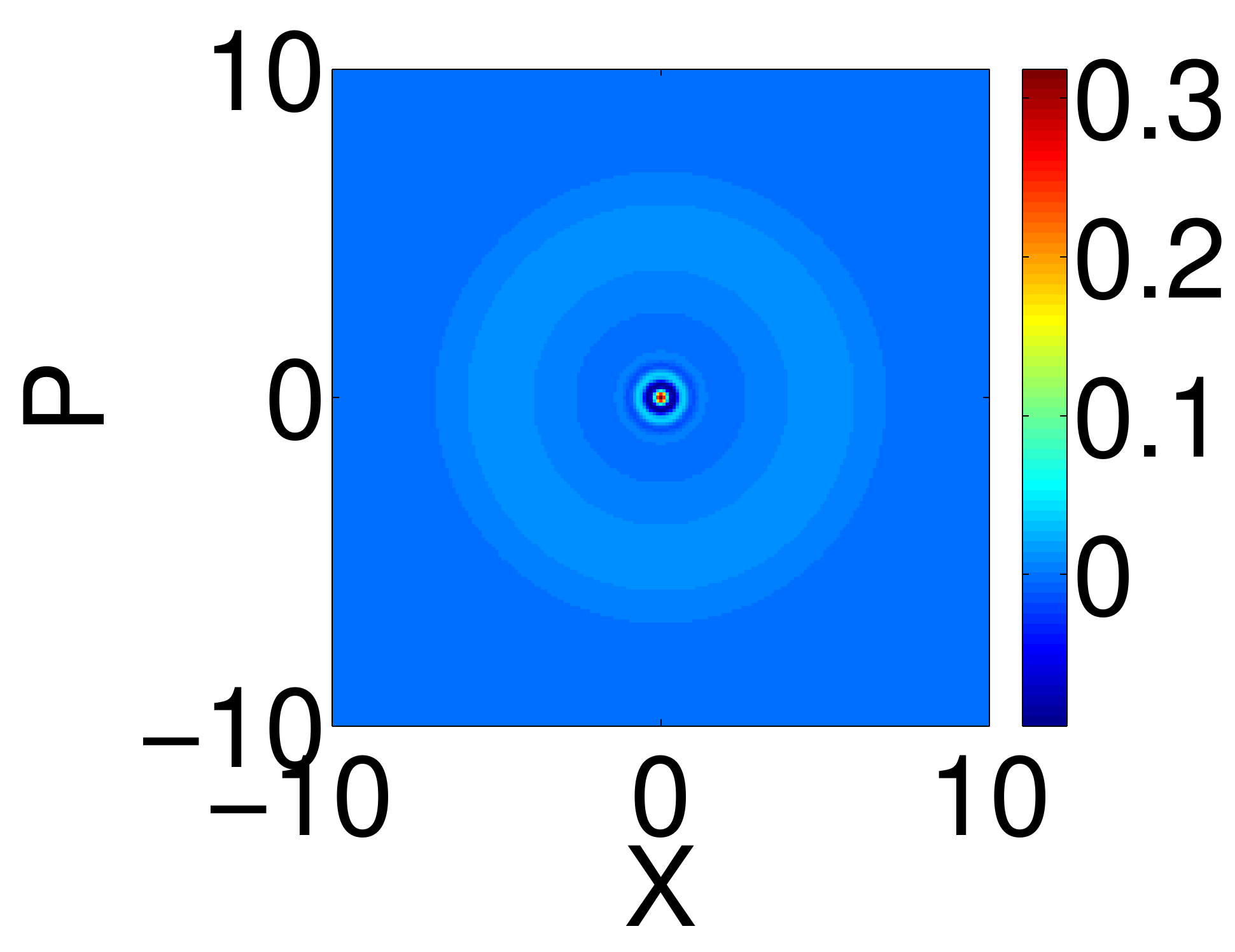}
    \caption{Fock, $t=8 \, \Gamma^{-1}$}
    \label{fig:WFock8}
       \end{subfigure}
    \caption[Nonlinear interaction: Wigner distribution function at various times]{Two--photon amplifier Wigner distribution functions at various times for an initial Poissonian--mixed state with $\bar{n}=4$ (a-c) and for an initial $n=4$ Fock state (d-f).}
       \label{fig:WFun}
\end{figure}

\begin{figure} [htb]
    \centering
    \begin{subfigure}[h] {0.12\textwidth}
        \includegraphics[width=\textwidth]
        {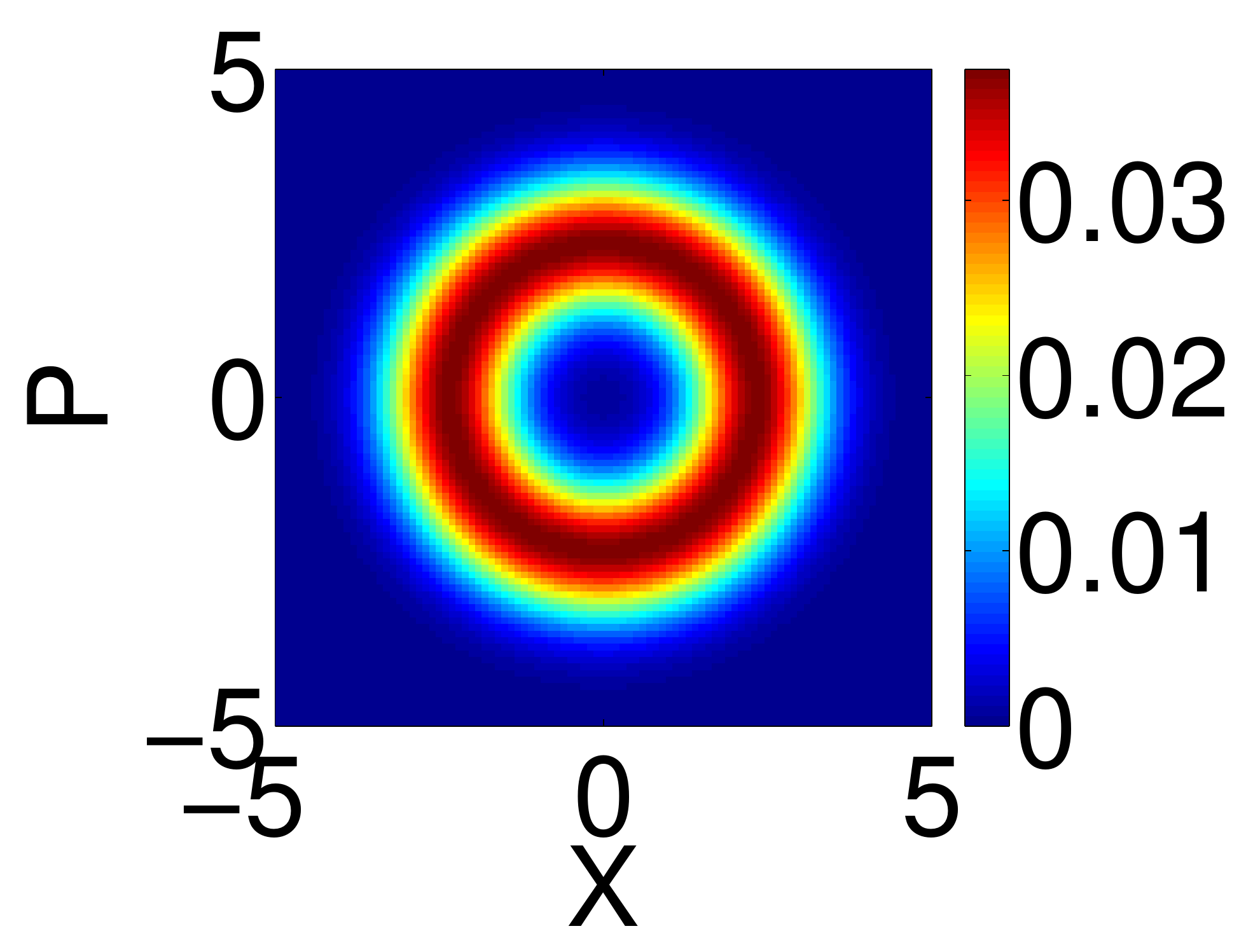}
    \caption{Mixed, $t=0$}
    \label{fig:WMixed30}
    \end{subfigure}
        \begin{subfigure}[h] {0.12\textwidth}
        \includegraphics[width=\textwidth]
        {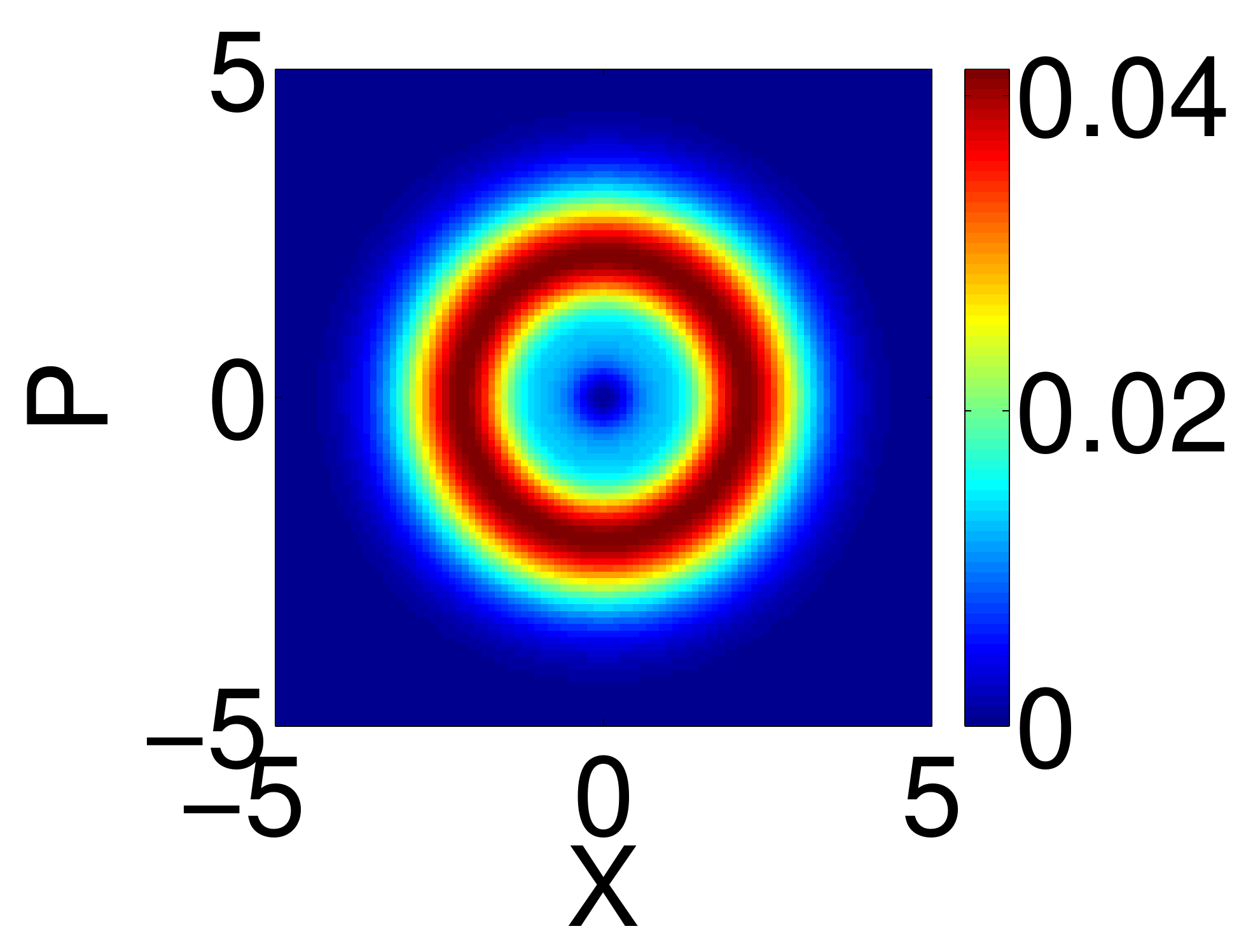}
    \caption{Mixed, $t=0.1 \, \Gamma^{-1}$}
    \label{fig:WMixed301}
    \end{subfigure}
        \begin{subfigure}[h] {0.12\textwidth}
        \includegraphics[width=\textwidth]
        {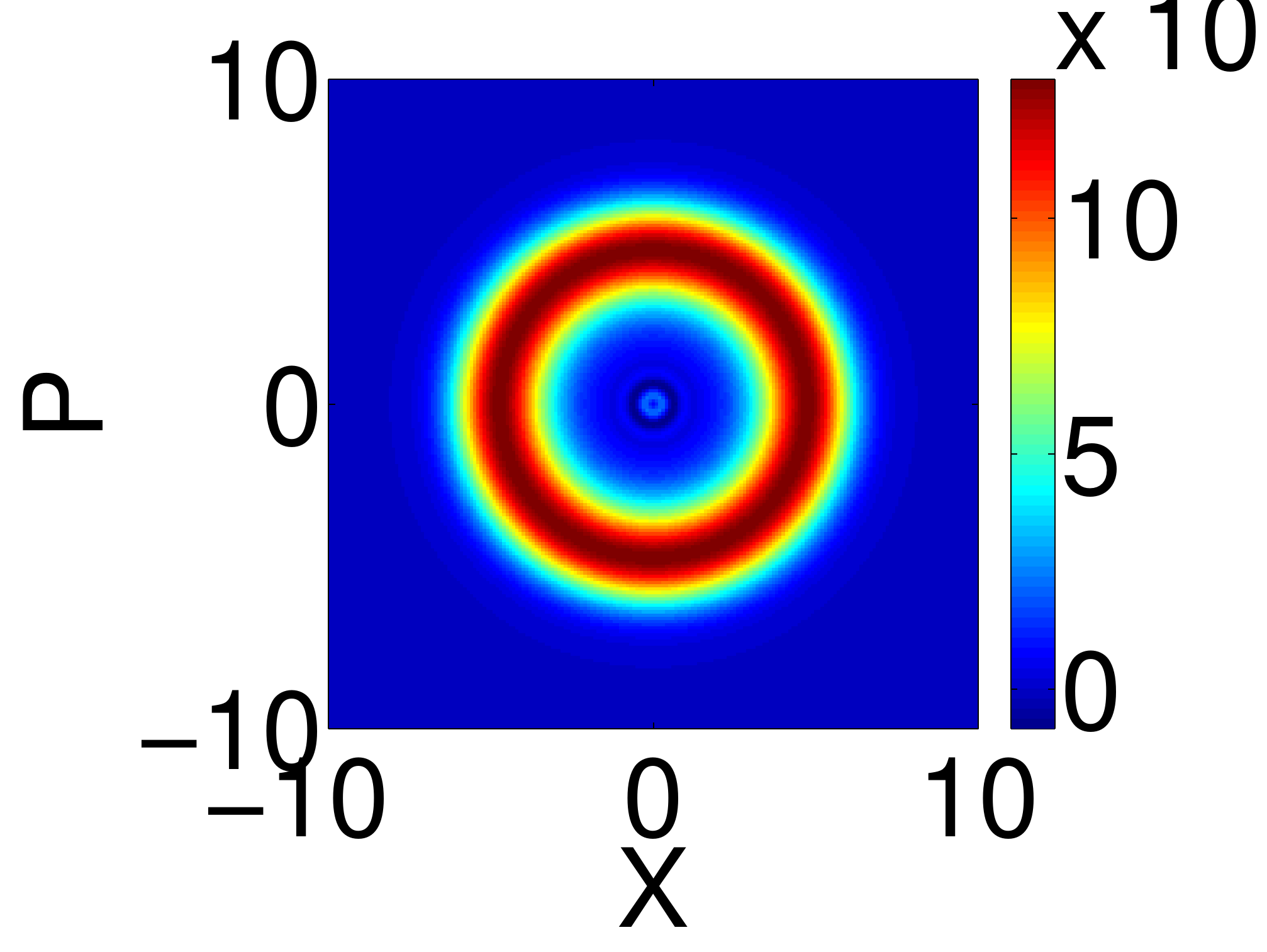}
    \caption{Mixed, $t=8 \, \Gamma^{-1}$}
    \label{fig:WMixed38}
       \end{subfigure}
          \begin{subfigure}[h] {0.12\textwidth} \vspace{-10pt}
        \includegraphics[width=\textwidth]
        {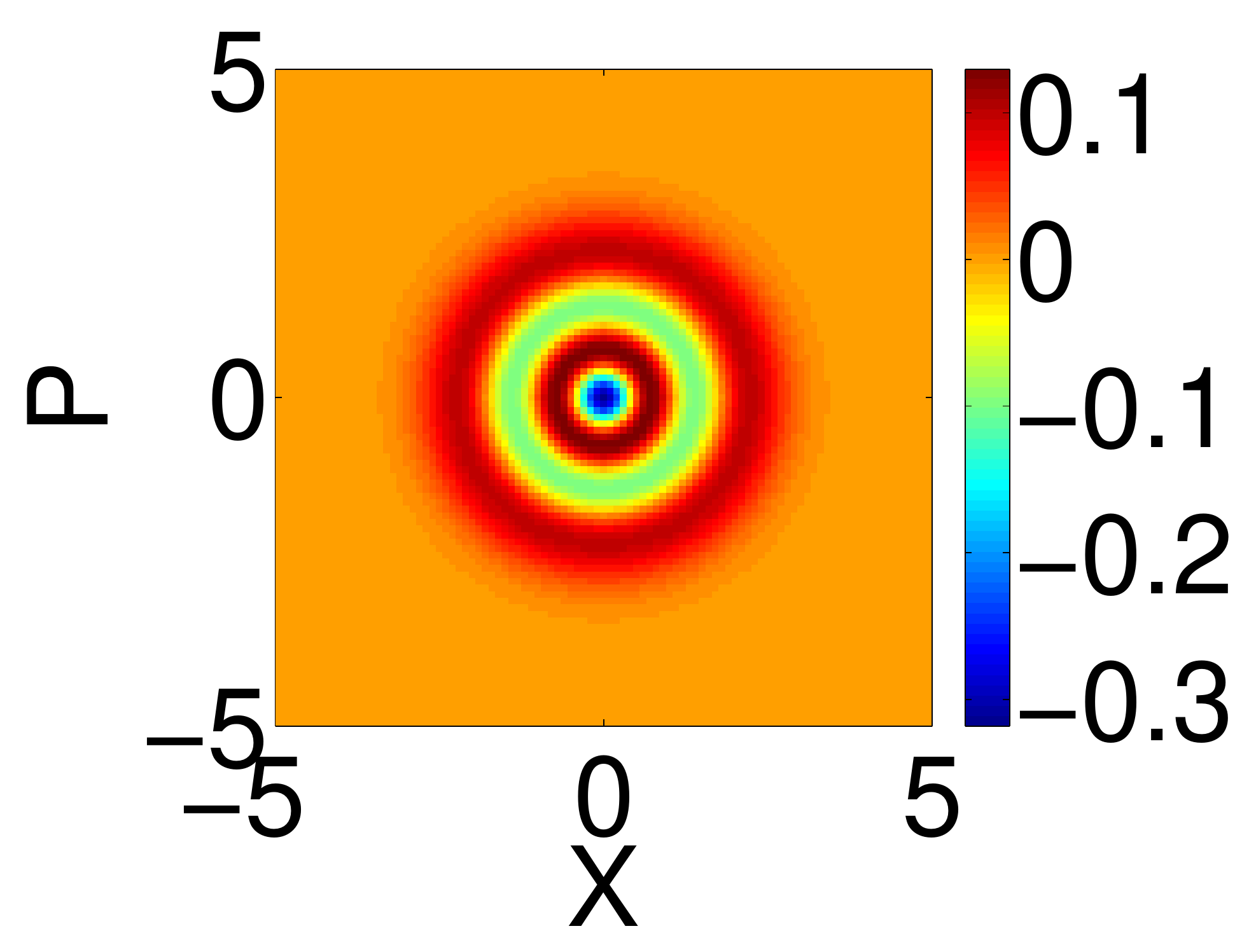}
    \caption{Fock, $t=0$}
    \label{fig:WFock30}
    \end{subfigure}
        \begin{subfigure}[h] {0.12\textwidth}
        \includegraphics[width=\textwidth]
        {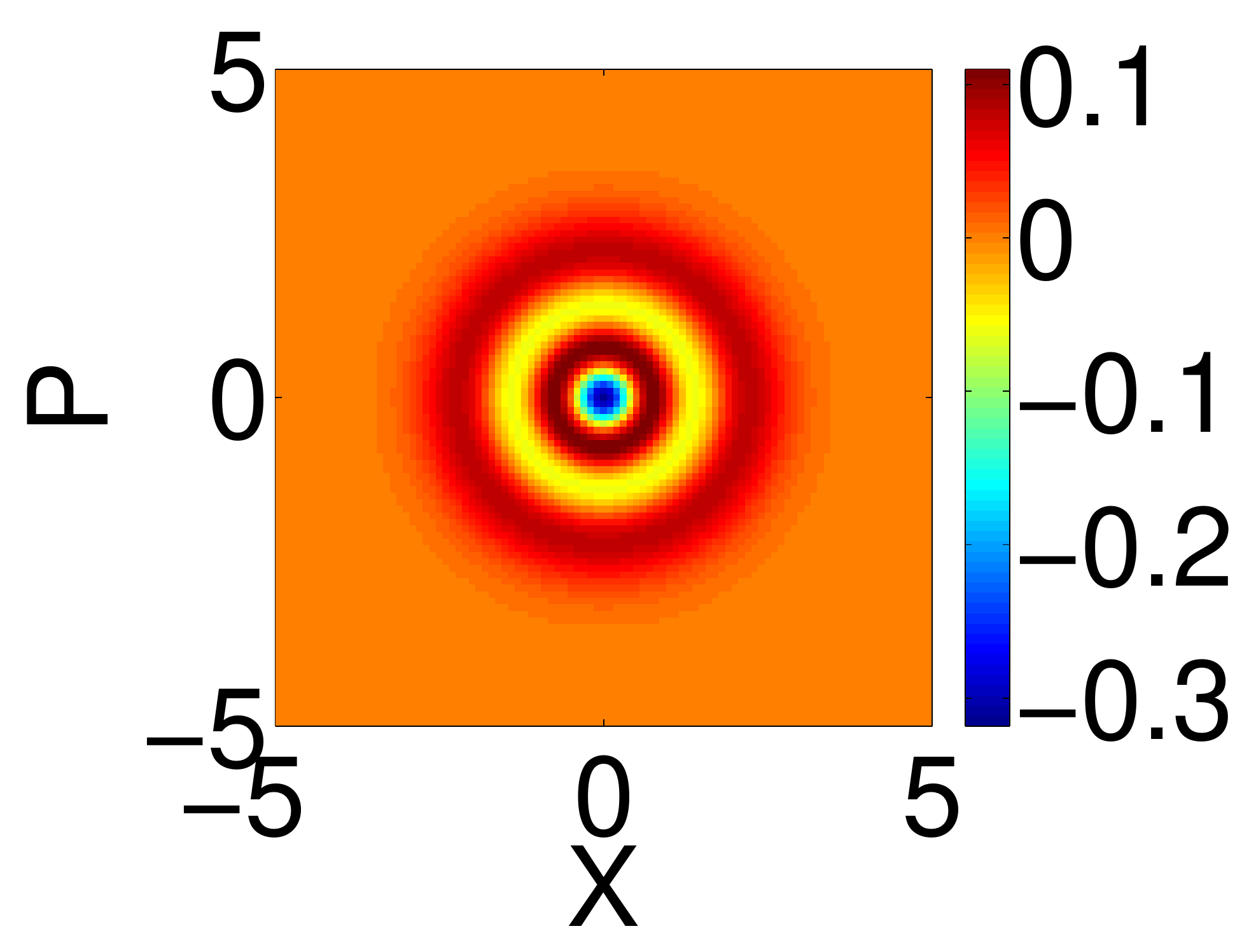}
    \caption{Fock, $t=0.1 \, \Gamma^{-1}$}
    \label{fig:WFock301}
    \end{subfigure}
        \begin{subfigure}[h] {0.12\textwidth}
        \includegraphics[width=\textwidth]
        {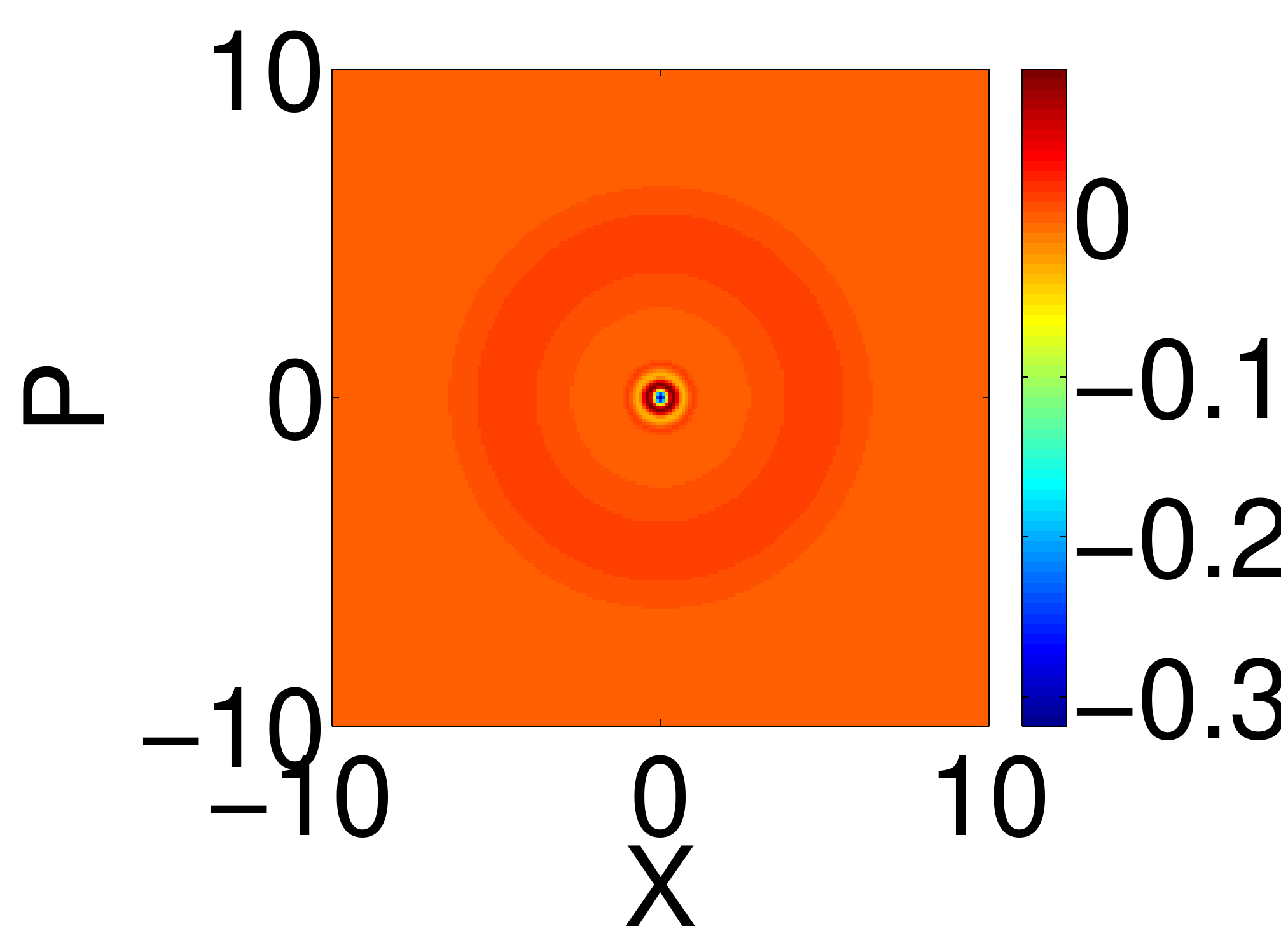}
    \caption{Fock, $t=8 \, \Gamma^{-1}$}
    \label{fig:WFock38}
       \end{subfigure}
    \caption[Nonlinear interaction: Wigner distribution function at various times]{Two--photon amplifier Wigner distribution functions at various times for an initial Poissonian--mixed state with $\bar{n}=3$ (a-c) and for an initial $n=3$ Fock state (d-f).}
       \label{fig:WFun3}
\end{figure}

Next, we plot the Wigner distribution function at different times for the two initial states discussed in Fig. \ref{fig:ComparePfEta} and Fig. \ref{fig:QFun}. We see in Fig.~\ref{fig:WFun} that the Wigner functions for the two states are completely different initially and at later times. Interestingly, negative amplitudes are clearly visible for the initial $n=4$ Fock state at thermodynamic times (when the atomic energy/entropy reaches a steady state). The effect of observing negative Wigner function amplitudes at extended times is dramatically pronounced for an initial odd Fock state, and is plotted for the initial $n=3$ Fock state in Fig. \ref{fig:WFun3}. The substantial negative amplitude renders its measurement in similar setups described in \cite{Haroche2002} according to the Lutterbach and Davidovich method \cite{Lutterbach_Davidovich1997}. Moreover, the position of the negative quasiprobability amplitudes in optical phase space at long times, is different when one compares odd (minimum at the origin) and even (minimum is spread evenly in rings around the origin) manifold states.

The fact that the two states are distinguishable after undergoing dissipation for very long times is a unique feature of the underlying dynamics. It can be understood by the following argument. The two--photon JCM Hamiltonian couples transitions $|n\rangle\rightarrow|n\pm 2\rangle$. In addition, the two dissipative super--operators are purely atomic in nature and hence they do not mix states in the Fock manifold. The natural outcome of this is that an initial even (odd) Fock state will evolve only in the even (odd) Fock manifold, and an initial mixed Poissonian state, which is in fact a probabilistic superposition of both even and odd manifolds, will evolve in the entire Fock space. The resulting outcome is that negative quaiprobability values of even Fock states will cancel with positive quaiprobability values of adjacent odd Fock states in the probabilistic superposition of the Poissonian--mixed state. For example, while an initial odd Fock state has a local negative minimum value at the origin, an even Fock state has a local positive maximum value at the origin. Therefore, the Wigner function serves as a fingerprint for a selected set of initial field states, at times where these states are both thermodynamically and Q--function indistinguishable.

We will now show that at long times, $t\geq 2\Gamma^{-1}$, the field density matrix is a probabilistic superposition of Fock states. We begin by noting that the full atomic--field bipartite density matrix assumes the following form after $t>2\Gamma^{-1}$:
\begin{equation}
\mbox{\boldmath$\rho_{af}$}=\left(
\begin{array}{c|c|c}
\mbox{\boldmath$P_{1,1}$} & \mbox{\boldmath$C$} &
\mbox{\boldmath$0$}\\\hline\mbox{\boldmath$C^{\dag}$} &
\mbox{\boldmath$P_{2,2}$} &
\mbox{\boldmath$0$}\\\hline\mbox{\boldmath$0$} & \mbox{\boldmath$0$}
& \mbox{\boldmath$P_{3,3}$}
\end{array}\right),\label{finalrhomf}
\end{equation}
where $\mbox{\boldmath$0$}$ is an $m\times m$ zero matrix,
$\mbox{\boldmath$P_{i,i}$}$ are diagonal matrices ($i=1,2,3$) whose
elements are $\mbox{\boldmath$\rho^{im,im}$}$, and $\mbox{\boldmath$C$}$ is an $m\times m$ correlation matrix whose only non-zero elements are the ones in the second diagonal above the main diagonal, $\mbox{\boldmath$\rho_{af}^{1m,2(m+2)}$}$. Since, topologically, tracing over the atom is equivalent to superimposing the three $\mbox{\boldmath$P_{i,i}$}$ matrices and summing them element by element, the resulting field density matrix is diagonal, and can be written as $\rho_{f}=\sum_{n}p_{n}|n\rangle\langle n|$. Therefore, at long times, the phase space functions of the amplified field state are a probabilistic superposition of individual Fock states phase space functions . For example:
\begin{equation}
\begin{split}
Q(\alpha)&\equiv\frac{1}{\pi}\bra{\alpha}\rho_f\ket{\alpha}=\\
&=\frac{1}{\pi}\bra{\alpha}\sum_{n}p_{n}|n\rangle\langle n|\ket{\alpha}=\sum_{n}p_{n}Q_{n}(\alpha),
\end{split}
\end{equation}
where $Q_{n}(\alpha)=\frac{1}{\pi}\bra{\alpha}n\rangle\langle n\ket{\alpha}$.

\begin{figure} [htb]
    \centering
    \begin{subfigure}[h] {0.12\textwidth}
        \includegraphics[width=\textwidth]
        {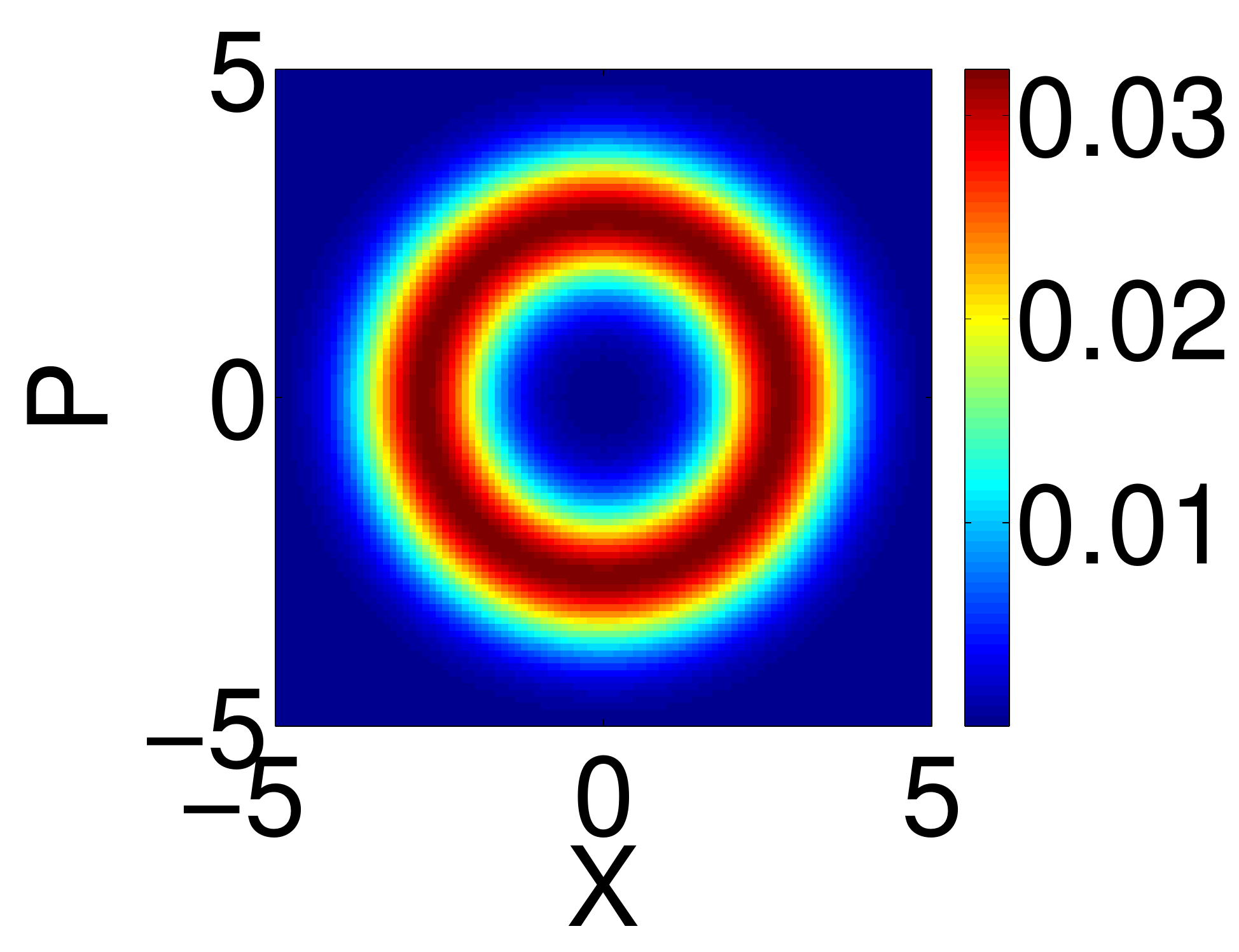}
    \caption{Mixed, $t=0$}
    \label{fig:WLinMixed0}
    \end{subfigure}
        \begin{subfigure}[h] {0.12\textwidth}
        \includegraphics[width=\textwidth]
        {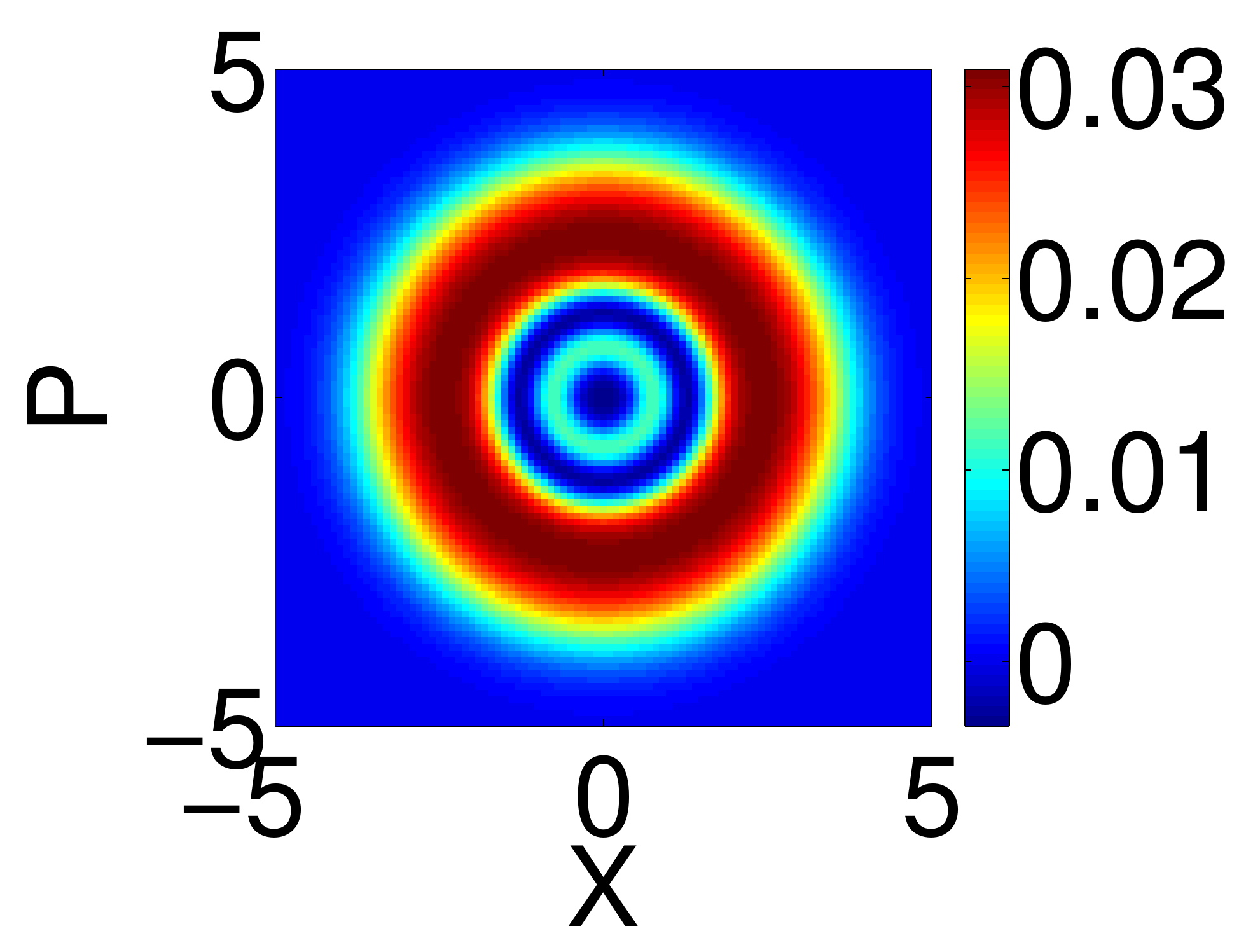}
    \caption{Mixed, $t=0.1 \, \Gamma^{-1}$}
    \label{fig:WLinMixed01}
    \end{subfigure}
        \begin{subfigure}[h] {0.12\textwidth}
        \includegraphics[width=\textwidth]
        {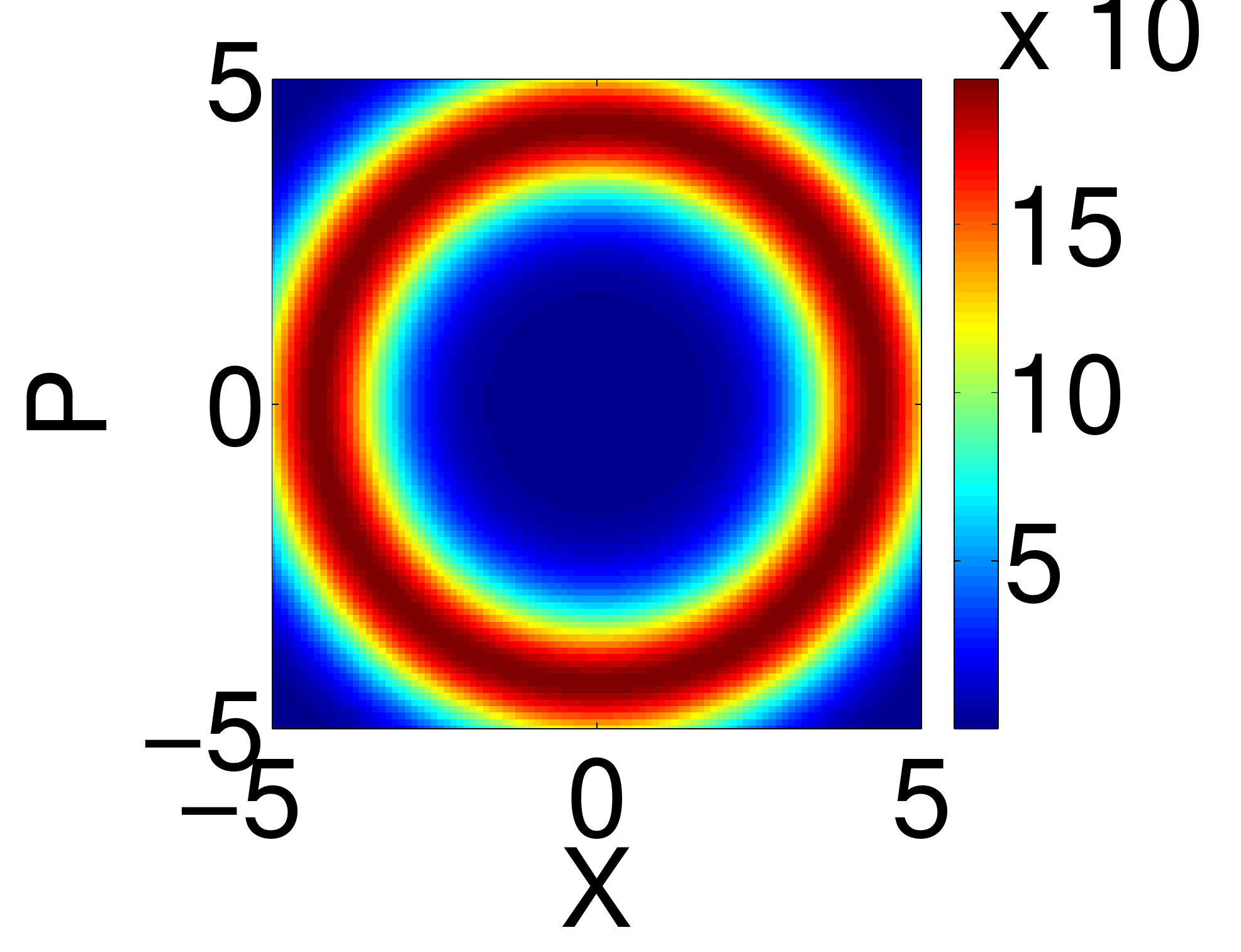}
    \caption{Mixed, $t=10 \, \Gamma^{-1}$}
    \label{fig:WLinMixed10}
       \end{subfigure}
          \begin{subfigure}[h] {0.12\textwidth}
          \vspace{-10pt}
        \includegraphics[width=\textwidth]
        {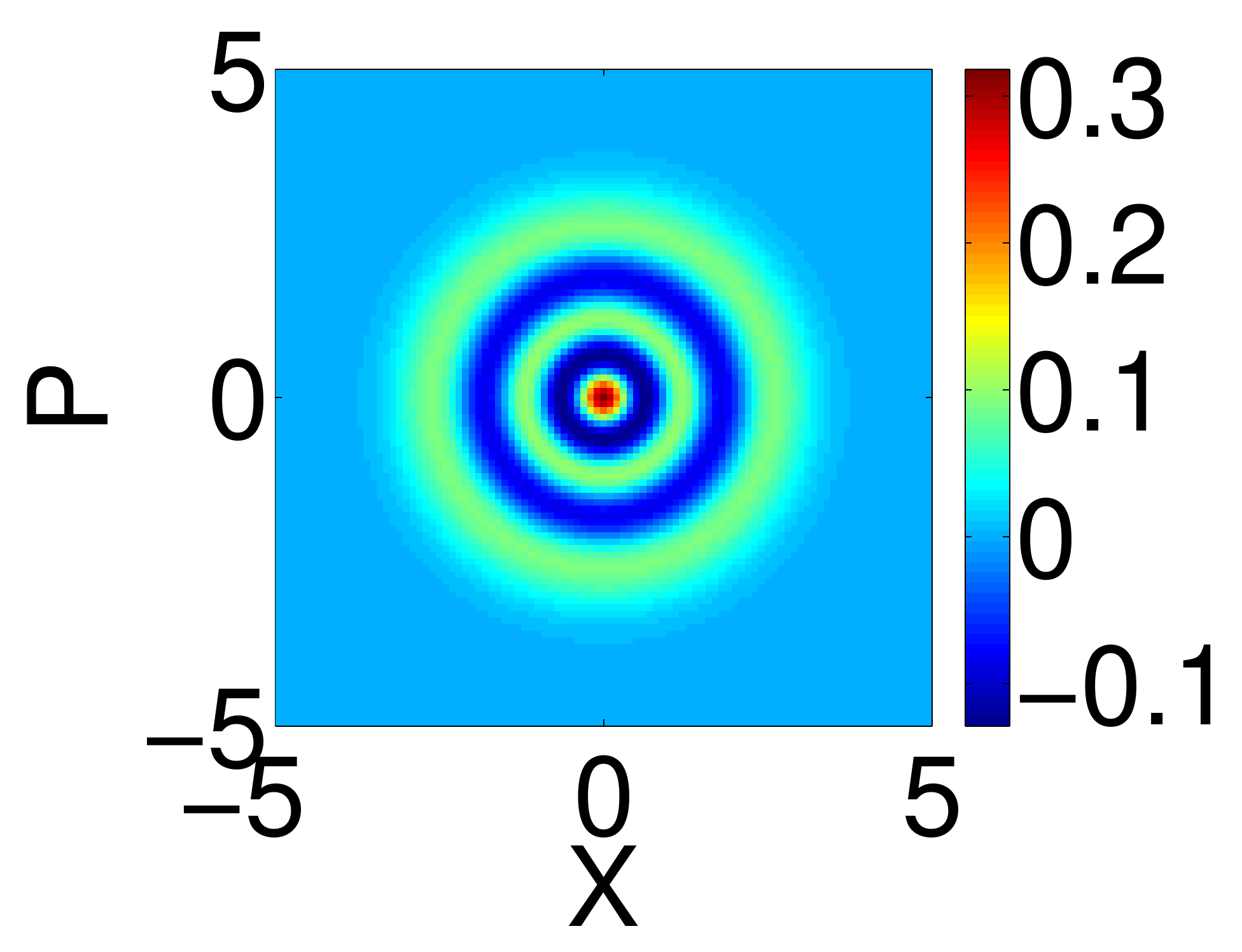}
    \caption{Fock, $t=0$}
    \label{fig:WLinFock0}
    \end{subfigure}
        \begin{subfigure}[h] {0.12\textwidth}
        \includegraphics[width=\textwidth]
        {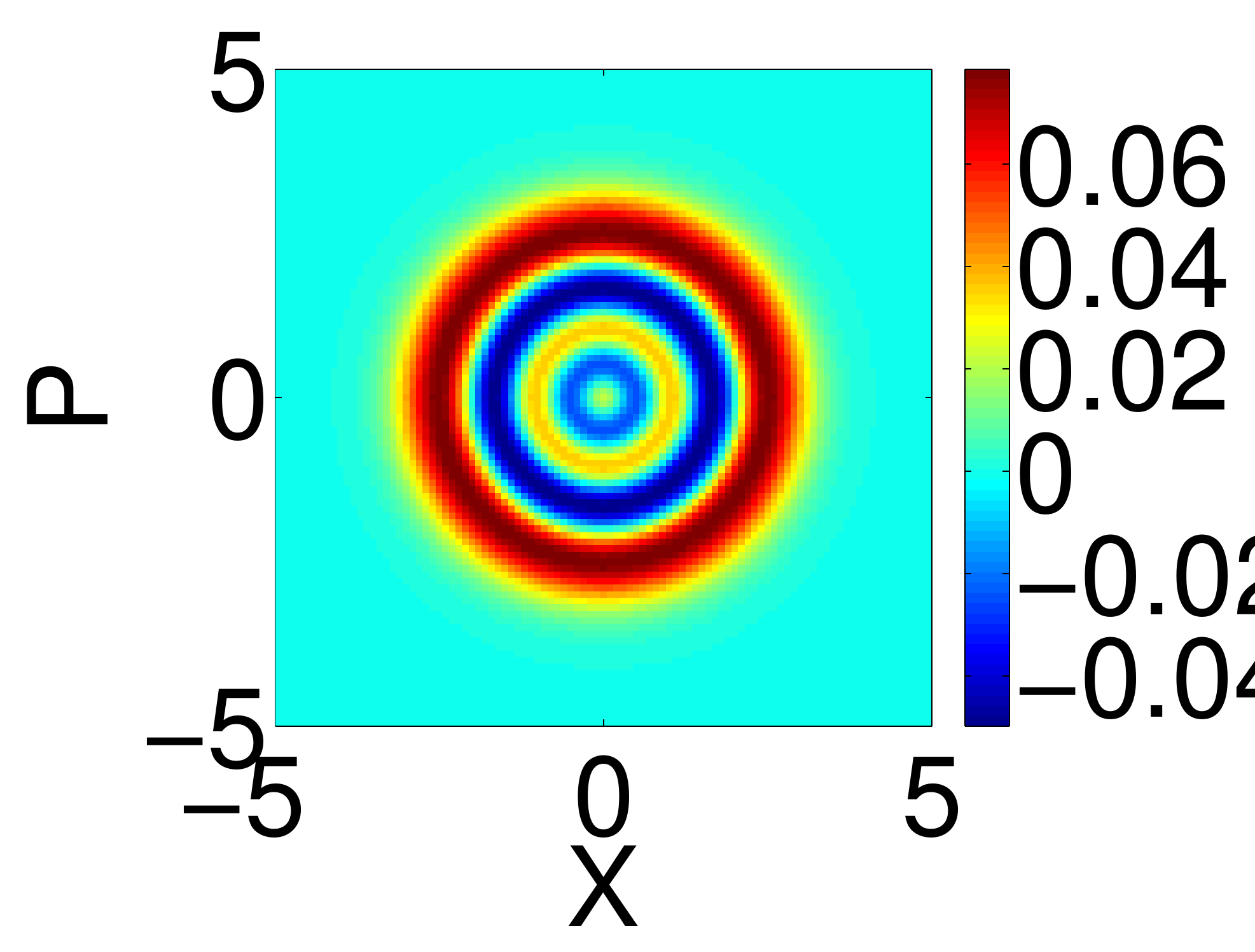}
    \caption{Fock, $t=0.1 \, \Gamma^{-1}$}
    \label{fig:WLinFock01}
    \end{subfigure}
        \begin{subfigure}[h] {0.12\textwidth}
        \includegraphics[width=\textwidth]
        {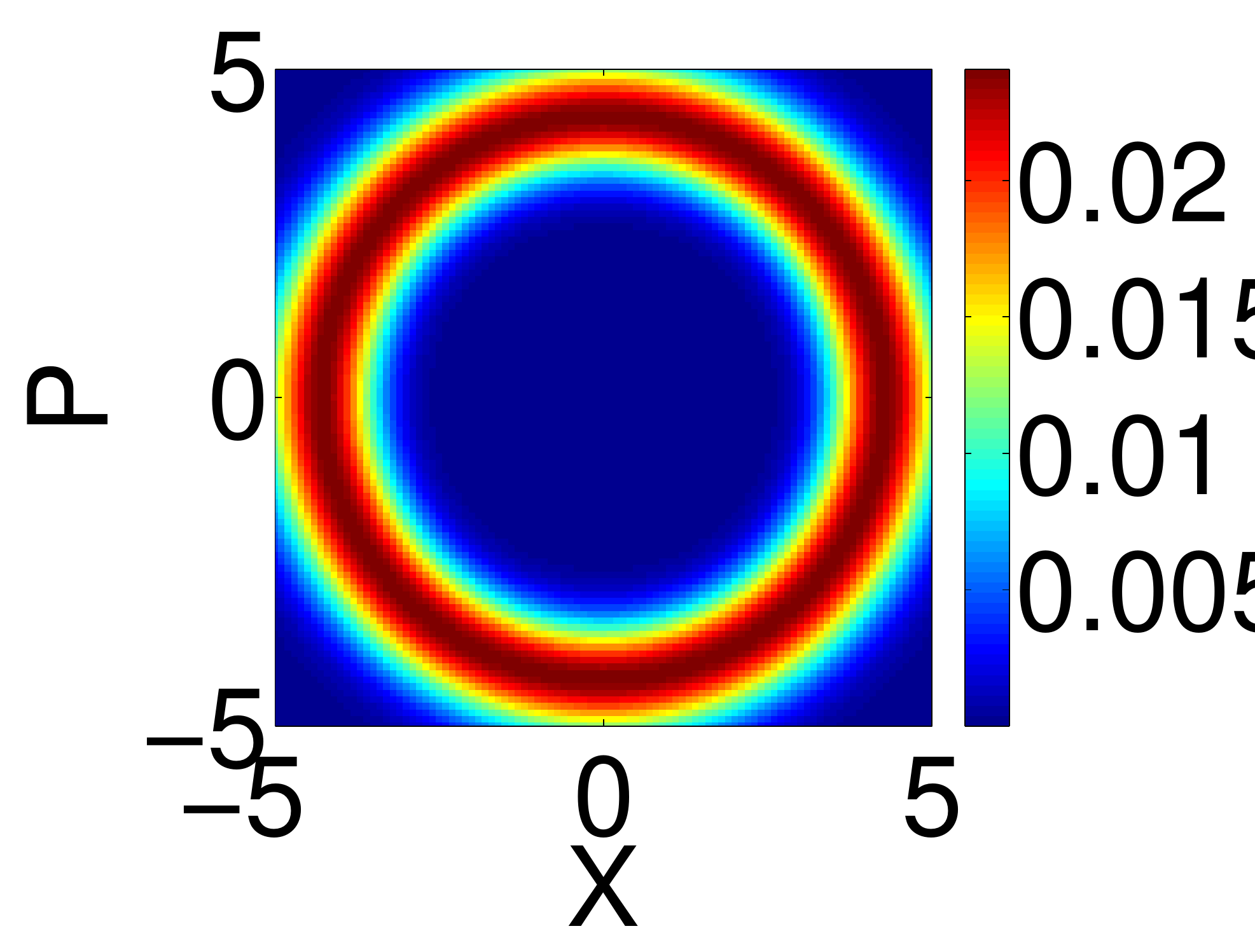}
    \caption{Fock, $t=10 \, \Gamma^{-1}$}
    \label{fig:WLinFock10}
       \end{subfigure}
    \caption[Linear interaction: Wigner distribution function at various times]{Linear interaction: Wigner distribution function at various times for an initial mixed state with Poisson distribution (a-c) and for an initial Fock state (d-f). See color-bars for numerical values.}
       \label{fig:WFunLin}
\end{figure}

To further emphasize that a non--linear amplifier may be associated with unique features in optical phase space, in Fig.~\ref{fig:WFunLin} we plot the Wigner functions for both the initial diagonal Poissonian (mixed) state with $\bar{n}=4$ and the $n=4$ Fock (pure) state, this time for a linear amplifier. Figure \ref{fig:WLinMixed0} to \ref{fig:WLinMixed10} represent the mixed initial state at different times, and Fig.~\ref{fig:WLinFock0} to \ref{fig:WLinFock10} represent the Fock initial state at different times. Since the interaction is linear, both even and odd states become populated at all times as more and more photons are created in the cavity, and hence the Wigner functions for the two initial states become very similar after $10 \, \Gamma^{-1}$. This is more pronounced when one plots the Q--function. Therefore, unlike the non--linear amplifier case, various initial amplified field states become both thermodynamically and phase space indistinguishable. In fact, measurement of the Wigner function  at long times for several initial field states, may resolve the type of amplification route involved (linear vs, non--linear).

\section{Conclusion\label{sec:Conc}}
We have presented both a thermodynamic and optical phase space analysis of a resonant non--linear (two--photon) light amplifier, with emphasis on long (thermodynamic) times.

Based on the thermodynamic formalism presented in \cite{Boukobza_Tannor2006b,Boukobza_Ritsch} we have established that a non--linear amplifier may be viewed as a heat engine. Moreover, we have demonstrated that different initial field states are thermodynamically equivalent. Their amplified power and efficiency are identical.

We have also derived an analytical formula for the efficiency of the resonant non--linear light amplifier, which is equal to the ratio of the resonance frequency and the pump frequency. Moreover, the efficiency formula is identical to that of the linear light amplifier. This efficiency formula was shown in \cite{Boukobza_Tannor2006a} to be less than the Carnot bound for a resonant quantum optical amplifier coupled to two bosonic reservoirs with identical coupling strengths to the working medium (the atom). This suggests that thermodynamic detailed balance at atomic--field resonance, keeping all coupling constants identical (atomic--field, aromic--reservoir), leads to a general efficiency formula bound by the Carnot limit regardless of the photonic cascade. This possible generalization for quantum optical amplifiers operating continuously as heat engines is reminiscent of the Curzon and Aahlborn \cite{Curzon_Ahlborn}, Chambdal \cite{Chambadal} and Novikov \cite{Novikov} general efficiency limit for endoreversible stroke engines with identical coupling strength between the working fluid and the two reservoirs as was noted by Uzdin and Kosloff \cite{kosloff2014c}.

The thermodynamic equivalence of individual initial light states at long times in resonant linear and non--linear light amplifiers, and the thermodynamic equivalence between the two types of amplifiers themselves raises the issue of distinguishability upon quantum measurement. At long times compared to the atomic--reservoir decay time, two different field states with an identical initial field excitation have very similar Q functions, making them also Q--function indistinguishable. However, their corresponding Wigner functions at long times are quite different. Furthermore, an initial Fock state which is amplified still shows negative Wigner amplitudes at extremely long times (thousands of Rabi cycles), while an initial mixed Poisson (superposition) state with similar initial excitation does not. Therefore, the prospect of measuring the Wigner function (as was suggested by Lutterbach and Davidovich \cite{Lutterbach_Davidovich1997} and as was measured by Haroche and coworkers \cite{Haroche2002}) offers an experimental tool in high Q cavities that both fingerprints the initial field state at long times, or a family of initial field states. Moreover, offers yet another manifestation of the discrete nature of matter--field interactions, also in the non--linear regime and with atomic dissipation present.

Finally, in view of recent work published in the field and of unique dynamical features presented here, we wish to point to several future research routes. The quantum optical heat engine discussed here and in \cite{Boukobza_Tannor2006a,Boukobza_Ritsch} is not a standard form of heat engine because it is not periodic. Its piston (the optical light field without losses) may be viewed as a piston that is always rising. Nevertheless, based on an effective dynamical temperature and the notion of non--passivity, one might calculate how much work (power) is extractable from such optical amplifiers, both at resonance and off--resonance. In addition, since negative Wigner amplitudes are a signature of quantum behavior, one might try and study their relation with atomic--field coherences and entanglement measures.

\section*{Acknowledgments}

\bibliography{MyBibRev}
\bibliographystyle{phaip}

\end{document}